\pdfoutput=1
\documentclass[12pt]{iopart}

\usepackage{latexsym} 
\usepackage{amssymb} 
\usepackage{amsfonts}
\usepackage{bm}
\usepackage{color}
\usepackage{times} 
\usepackage{units}
\usepackage{hyperref}

\usepackage[utf8x]{inputenc}
\usepackage{graphicx} 
\usepackage[squaren]{SIunits} 

\graphicspath{ {./plots/} }

\newcommand{\Msol}{\ensuremath{M_\odot}} 
\newcommand{\Gauss}{\ensuremath{\mathrm{G}}}

\newcommand{\parsec}{\mathrm{pc}} 

\newcommand{\eg}{e.g.,~}

\begin{document}


\title[GRMHD Simulations of BNS Mergers with the APR4 EOS]{General
  Relativistic Magnetohydrodynamic Simulations of Binary Neutron Star
  Mergers with the APR4 Equation of State}

\author{A Endrizzi, R Ciolfi, B Giacomazzo, W Kastaun, and T Kawamura}

\address{Physics Department, University of Trento, 
via Sommarive 14, I-38123 Trento, Italy}

\address{INFN-TIFPA, Trento Institute for Fundamental Physics
  and Applications, via Sommarive 14, I-38123 Trento, Italy}

\eads{\mailto{andrea.endrizzi.2@unitn.it},
  \mailto{bruno.giacomazzo@unitn.it}}


\begin{abstract}
We present new results of fully general relativistic
magnetohydrodynamic (GRMHD) simulations of binary neutron star (BNS)
mergers performed with the Whisky code. All the models use a piecewise
polytropic approximation of the APR4 equation of state (EOS) for cold
matter, together with a ``hybrid" part to incorporate thermal effects
during the evolution. We consider both equal and unequal-mass models,
with total masses such that either a supramassive NS or a black hole
(BH) is formed after merger. Each model is evolved with and without a
magnetic field initially confined to the stellar interior. We present
the different gravitational wave (GW) signals as well as a detailed
description of the matter dynamics (magnetic field evolution, ejected
mass, post-merger remnant/disk properties). Our simulations provide
new insights into BNS mergers, the associated GW emission and 
the possible connection with the engine of short
gamma-ray bursts (both in the ``standard'' and in the
``time-reversal'' scenarios) and other electromagnetic
counterparts.
\end{abstract}

\noindent Online supplementary data available from \url{stacks.iop.org/cqg/33/164001/mmedia}



\submitto{\CQG}

\maketitle

\section{Introduction}
\label{sec:intro}

After the recent detection \cite{LIGO:BBHGW:2016} of gravitational waves (GW) 
from the merger of two black holes (BHs) by the ground-based LIGO interferometers, 
it seems to be only a matter of time until GWs from merging neutron stars (NS) 
are detected as well (see~\cite{Abadie2010} for estimated event rates). 
BNS mergers may also power bright electromagnetic
(EM) signals, including short gamma-ray bursts (SGRBs,
see~\cite{Berger2014} for a review). The imminent 
integration of the Virgo detector into the working GW network will
improve the sky localization and enlarge the chances of detecting
EM counterparts in the follow-up of a GW detection. 
Furthermore, the matter ejected
during BNS mergers is thought to be responsible, at least partially, 
for the creation of heavy elements in the universe (see 
\eg\cite{Mendoza:2015:055805, Korobkin:2012:01112012}).

BNS systems can be divided into two main categories: ``high-mass''
BNSs that form a BH or a hypermassive NS (HMNS) after merger, and
``low-mass'' BNSs that form instead a long-lived supramassive NS
(SMNS) or even a stable NS.  An HMNS is an NS whose mass is above the
maximum mass for an uniformly rotating NS (hypermassive limit) and
that will collapse to a BH in less than ${\sim}1$ second after
merger. The BH formed by ``high-mass'' BNS mergers may be surrounded
by an accretion disk, which is thought to be a necessary 
(however not sufficient) condition to power a relativistic 
jet and produce an SGRB. Preliminary simulations performed with the 
{\tt Whisky} code showed the possible formation of a strongly collimated 
magnetic field along the BH spin axis~\cite{Rezzolla:2011:6}, but a different
simulation from another group was not able to produce similar
results~\cite{Kiuchi:2014:41502}. The only evidence of jet formation,
up to now, has been provided very recently for an equal-mass NS-NS
merger~\cite{Ruiz2016}, following a similar result for an NS-BH
merger~\cite{Paschalidis:2015:14}.

An SMNS is an NS with mass above the maximum mass for a
non-rotating NS, but below the hypermassive limit. In this case, uniform
rotation can be sufficient to support the star against the
collapse to a BH. However, on a spin-down timescale (minutes to hours) the
star will eventually collapse. Therefore, the merger of a
``low-mass'' system results in a NS that either collapses on a very
long time scale, or does not collapse at all. 
The observations of 
${\sim}2 \usk M_{\odot}$ NSs~\cite{Demorest:2010:1081, Antoniadis:2013:448} 
support the idea that at least a significant fraction of BNS mergers will lead to the
formation of SMNSs or even stable NSs. The possibility of forming
highly magnetized NSs of this kind was also shown in recent
simulations \cite{Giacomazzo2013ApJ...771L..26G, Giacomazzo:2015,Kiuchi:2015:1509.09205}.
Long-lived remnant NSs are important in the context of the 
``magnetar''~\cite{Zhang2001, Metzger2008} and 
``time-reversal''~\cite{Ciolfi:2015:36,Ciolfi:2015:PoS}
scenarios for SGRBs (see \cite{Rezzolla2015} for an alternative proposal).
These scenarios are also supported by the observation of long-lasting X-ray
plateaus in the afterglow emission of many SGRBs~\cite{Rowlinson2013}.  

Lifetime estimates of SMNSs based only on the total mass are
necessarily very broad. More detailed
models would have to take into account the exact rotation profile of 
the remnant and its evolution. 
A frequently used assumption in models of merger remnants is the so 
called j-constant law, featuring a rapidly rotating core and a slower
rotation in the outer layers of the star. However, a recent study 
\cite{Kastaun:2015:064027} of the rotation profile in merger remnants 
found completely different rotation profiles, with a slowly rotating
core and faster rotating outer layers. Further investigation of this
important aspect is needed. 

In order to study the different scenarios, it is necessary to perform
fully general relativistic magneto-hydrodynamic (GRMHD) simulations of
both ``high-mass'' and ``low-mass'' magnetized BNS mergers.  Since the
NS equation of state (EOS) is still largely unknown (in particular for
the high-density region in the NS core) it is important to explore
different models. In GRMHD there have been very few publications
considering different EOSs. Most of them have considered a simple
ideal-fluid EOS~\cite{Anderson2008PhRvL.100s1101A, Liu2008,
Giacomazzo2009, Giacomazzo2011PhRvD..83d4014G, Rezzolla:2011:6,
Giacomazzo2013ApJ...771L..26G, Giacomazzo:2015, Ruiz2016}, very few
piece-wise polytropic approximations~\cite{Kiuchi:2014:41502,
Kiuchi:2015:1509.09205}, and only one a finite temperature tabulated
EOS including neutrinos~\cite{Palenzuela2015}. Moreover, all the
simulations have considered only equal-mass magnetized BNSs
(except~\cite{Liu2008}, using however a simple ideal-fluid EOS).

In this paper we present our new set of GRMHD simulations describing
both equal and unequal-mass models of magnetized BNSs. We considered
both an ``high-mass'' system, that collapses promptly to a BH after
merger, and ``low-mass'' systems that produce SMNSs. We used a
piecewise polytropic approximation of the APR4 EOS including also
thermal effects. In all cases we studied the impact of magnetic field
evolution on the dynamics, the GW emission, formation of disks, and
the possible connection with EM emission.

Our paper is organized as follows. \Sref{sec:numerics} describes
the initial data and the numerical methods used to evolve them. In
\sref{sec:results} we describe the general dynamics of these systems,
the structure of the disks that are formed, the rotation profile of
the merger remnants, the evolution of the magnetic field, the GW
signals, and the ejecta. In \sref{sec:conclusions} we summarize
our main results. Throughout the paper we use geometric units with
$c=G=M_{\odot}=1$, unless specified otherwise.
Baryonic mass is defined as baryon number times a formal baryon mass of 
$1.66 \times 10^{-24} \usk\gram$.


\section{Setup}
\label{sec:numerics}

\subsection{Numerical Methods}
\label{sec:methods}

All the numerical simulations presented in this paper were carried out
using the publicly available Einstein
toolkit~\cite{Loeffler:2012:115001} (``Wheeler'' release) combined
with our fully GRMHD code \texttt{Whisky}~\cite{Giacomazzo:2007:235,
Giacomazzo2011PhRvD..83d4014G, Giacomazzo2013ApJ...771L..26G}. The
GRMHD equations are written in a flux-conservative form using the
``Valencia'' formulation~\cite{Anton:2005gi} and then solved using
high-resolution shock capturing methods. In particular, the fluxes are
computed via the standard HLLE formula~\cite{Harten:1983:35} at the
boundaries between cells where primitive variables are reconstructed
via the PPM scheme~\cite{Colella:1984:174}. As in all GRMHD
simulations, we enforce a positive rest-mass density $\rho$ by
imposing an artificial atmosphere with fixed density and zero
3-velocity. Our choice for the atmosphere density is
$\rho_a=6.2\times10^{6}\usk\gram\per\centi\meter\cubed$, which is one
order of magnitude lower than what was used in our previous GRMHD
simulations~\cite{Giacomazzo:2007:235, Giacomazzo2011PhRvD..83d4014G,
Rezzolla:2011:6, Giacomazzo2013ApJ...771L..26G}. When a BH is formed,
hydro variables are excised (set to the artificial atmosphere) inside
a region bounded by the apparent horizon scaled down by a factor of
0.6.  This is done in order to avoid failures during the conversion
from conserved to primitive variables (see~\cite{Giacomazzo:2007:235}
for details).

In order to preserve the divergence-free character of the magnetic
field, we evolve directly the vector potential using the modified
Lorenz gauge~\cite{Etienne:2011re,Farris2012}. This guarantees
divergence-free magnetic fields and avoids spurious magnetic field
amplifications at the boundary between refinement levels. Adaptive
mesh refinement is implemented via the Carpet driver which is part of
the Einstein toolkit. In all the simulations we employed $6$
refinement levels, with a resolution of $0.15\Msol\approx222\usk\meter$ 
for the finest level. During inspiral, the two finest levels follow 
the NSs, which are completely contained in the finest grid.
Shortly before merger, the moving grids are replaced by larger fixed
grids. The smallest covers a radius of $30\usk\kilo\meter$,  sufficient 
to contain the post-merger remnant. The outer boundary is located at 
$794\usk\kilo\meter$. In order to save computational resources, 
we apply reflection symmetry with respect to the equatorial plane. 
The Einstein equations are solved via the
BSSNOK~\cite{Baumgarte:1998:24007,Shibata:1995:5428,Nakamura:1987:1}
formalism using the MacLachlan code, which is also part of the Einstein
toolkit.

Our initial models are built using the publicly available LORENE
library~\cite{Gourgoulhon:2001:64029}. LORENE is a multi-domain
spectral code that computes the initial data assuming a quasicircular
orbit, an irrotational fluid-velocity field, and a conformally flat
spatial metric.



\begin{figure}[t]
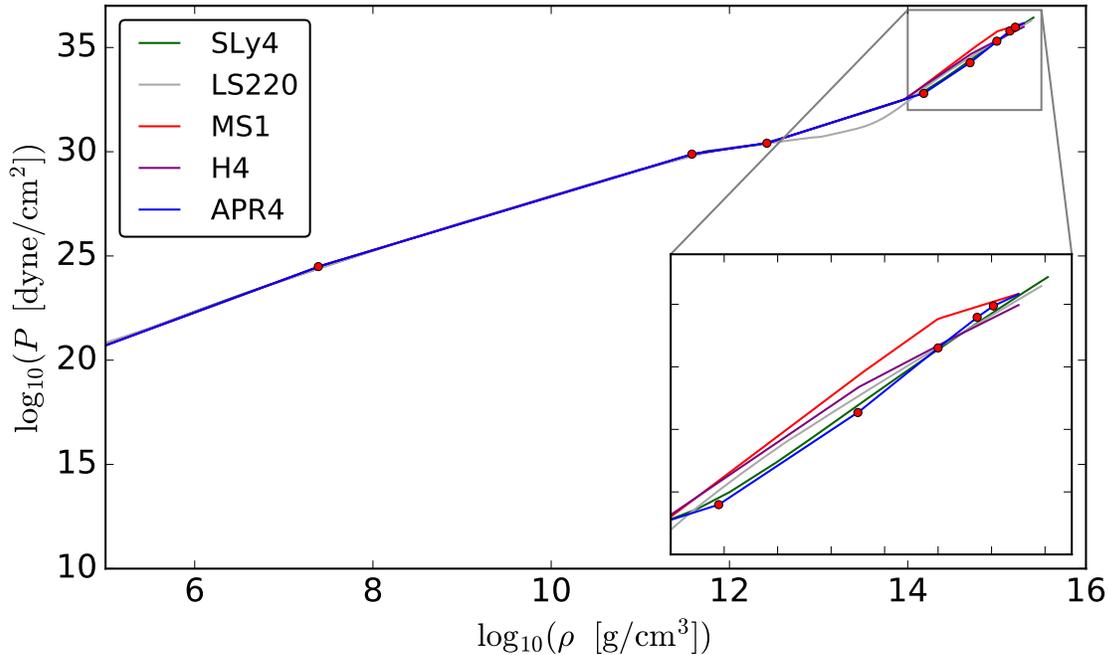

  \centering
  \includegraphics[width=0.99\columnwidth]{{{EOS_comparison}}}
  \caption{Pressure versus rest mass density relation for the
    piecewise polytropic approximation of the APR4 EOS used in this
    work. For comparison, we also show some other EOSs.}
  \label{fig:EOS_p2rho}
\end{figure}

\subsection{Initial Data}
\label{sec:init_data}

All the initial data in this work employ a piecewise polytropic
approximation of the APR4 EOS introduced by \cite{Akmal:1998:1804}.
The parameters for the polytropic segments are taken from
\cite{Read:2009:124032}.  For the evolution, we add a thermal
component to obtain a hybrid EOS given by
\begin{equation}
P\left(\rho, \epsilon\right) = 
  P_\text{cold}\left(\rho\right) 
  + \left(\Gamma_\text{th} - 1 \right) \left(\epsilon - \epsilon_\text{cold}\left(\rho\right) \right) \rho
\end{equation}
where we choose $\Gamma_\text{th} = 1.8$ (see discussion in 
\cite{Bauswein:2010:84043}). Further, we noticed that
the piecewise polytropic approximation \cite{Read:2009:124032} of the
APR4 EOS is only causal up to a density of $1.45 \times 10^{15}
\usk\gram\per\centi\meter\cubed$, above which the sound speed becomes
superluminal.  The critical density is larger than the central density
of all the NSs used for our initial data.  During the evolution
however, the density can exceed this value, either during a short
period when the stars are merging or while undergoing collapse to a
BH.  We therefore add two more high-density pieces, one with $\Gamma =
3$ and starting at density $1.4 \times 10^{15}
\usk\gram\per\centi\meter\cubed$, and one with $\Gamma = 2$ for
densities above $1.61 \times 10^{15}
\usk\gram\per\centi\meter\cubed$. The resulting hybrid EOS is fully
causal (regardless of temperature), although it is probably not
particularly realistic in the high density part.

\begin{figure}[t]
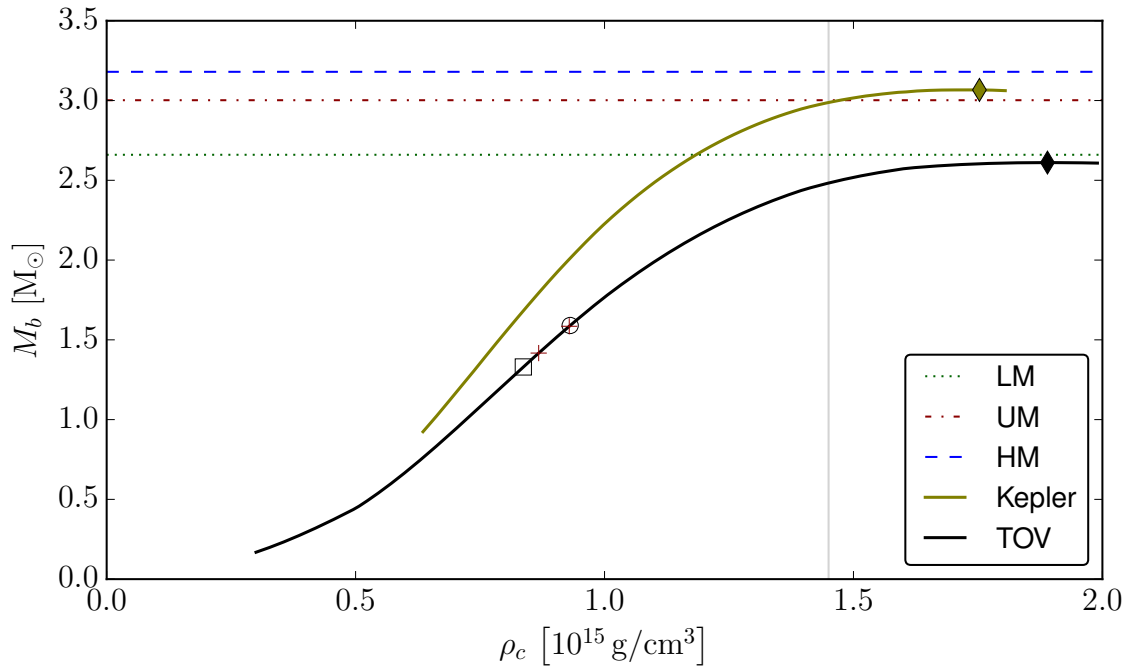

   \centering
   \includegraphics[width=0.99\columnwidth]{{{init_data}}}
   \caption{Total baryonic mass as function of the central rest mass
     density, for nonrotating NSs and for uniformly rotating NSs at
     the mass-shedding limit, employing the EOS used for all our
     initial data.  The horizontal lines correspond to the total
     baryonic masses of the three models we evolved. The individual
     stars of the low-, unequal-, and high-mass binaries are marked by
     square, plus, and circle markers, respectively. The diamond
     symbols denote maximum mass models.  The vertical line marks the
     density where the original APR4 approximation given in
     \cite{Read:2009:124032} becomes non-causal and had to be
     modified.}
   \label{fig:init_data}
\end{figure}

\Fref{fig:EOS_p2rho} shows pressure versus rest mass density for the
cold part in comparison to other well known EOSs. They differ only at 
high densities, since the EOS for the density range of the NS crust is 
better constrained by current understanding of nuclear physics.
We should note that the same low density EOS is used together with the
added thermal part for the evolution of ejected matter. Since such
matter is shock-heated, we expect the thermal part to dominate in this
regime.
We computed sequences of TOV stars as well as uniformly rotating stars
with maximal rotation using the piecewise polytropic APR4 EOS.
\Fref{fig:init_data} shows the baryonic mass versus the central
density for these sequences.  
We find that the supramassive mass range lies between
$2.61$--$3.07 \usk M_\odot$.

We evolve three different initial models, which are summarized in
\tref{tab:init_param}.  The first (``high mass'', HM) is an equal mass 
model with total mass in the hypermassive range (cf.
\fref{fig:init_data}), which can either form a metastable HMNS or
directly collapse to a BH.  
The second equal mass model is in the supramassive mass range and
expected to form a long-lived remnant (``low mass'', LM). Our third
model is an unequal mass binary with mass ratio 0.905 (``unequal
mass'', UM). Although its total mass is in the 
upper supramassive regime, the resulting remnant can be somewhat
lighter since unequal mass models typically form more massive disks
during merger.

Each of the three models is evolved with and without an initial
magnetic field.  Since the LORENE code cannot yet compute magnetized
BNS models, we manually add a poloidal magnetic field using a simple
analytic prescription for the vector potential:
\begin{equation}
\label{eq:Avec}
A_{\phi} \equiv \varpi^2 A_b\, {\rm max}\,(p-p_{\rm cut},0)^{n_{\rm s}} \,,
\end{equation}
where $\varpi$ is the coordinate distance to the NS axis (orthogonal to 
the orbital plane). The field is confined to the NSs, using a cutoff 
pressure $p_{\rm cut}=0.04$ of the maximum (central) pressure. The 
exponent $n_{\rm s}=2$ determines the degree of differentiability of 
the potential~\cite{Giacomazzo2011PhRvD..83d4014G}. The strength of 
the field, determined by $A_b$, is chosen such that the maximum field
strength is $1.0\times 10^{13}\usk\Gauss$. For the unequal mass model,
this is done separately for each star.  The corresponding magnetic
energy (see \tref{tab:init_param}) is below $10^{-11}$ of the NS
binding energy. Hence we can neglect the impact on the hydrostatic
equilibrium, and also the violation of the general relativistic
constraints. We stress however that finding a stable magnetic field 
configuration for NSs is still an unsolved problem. The prescribed 
magnetic field topology will decay into an unordered field during 
the inspiral.

\begin{table}[t]
	\caption{Initial data parameters. $M_{b}$ is the total
          baryonic mass of the systems, $M_g$ is the gravitational
          mass of each star at infinite separation, and
          $q=M_g^1/M_g^2$ the mass ratio. $M_g/R_c$ is the compactness
          (dimensionless).  $f_0$ and $d$ denote initial orbital
          frequency and proper separation, respectively.  $E_{B}$ is
          the initial magnetic energy of the magnetized models, which
          are otherwise identical to the non-magnetic ones.}
	
	\begin{indented}\item[]\begin{tabular}{@{}llll}
	\br 
	Model & HM & LM & UM \\\mr
	$q$                  & $1$     & $1$     & $0.905$ \\
	$M_{b}$ [M$_\odot$]  & $3.18$  & $2.66$  & $3.01$ \\
	$M_{g}$ [M$_\odot$]  & $1.43$  & $1.22$  & $1.29,1.42$ \\
	$M_g/R_c$            & $0.186$ & $0.159$ & $0.168, 0.186$ \\
	$f_0$ [Hz]           & $288$   & $270$   & $282$ \\
	$d$ [km]             & $60.0$  & $57.5$  & $59.0$ \\
	$E_{B}$ [$10^{42}$erg] &$1.58$ & $1.52$  & $1.55$ \\\br 
	\end{tabular}\end{indented}
	\label{tab:init_param}
\end{table}


\section{Results}
\label{sec:results}
In the following, we present the outcome of our simulations. 
Note that the results for the post merger phase lack reliable
error estimates since convergence could only be demonstrated 
until merger. A detailed discussion of the numerical errors 
can be found in \ref{sec:app}.

\subsection{General Dynamics}
\label{sec:gen_dim}

In the following, we provide an overview of the evolution of the three
models.  We will focus on the non-magnetic case. The influence of the
magnetic field is very small and will be discussed in later sections.

\begin{figure}[t]
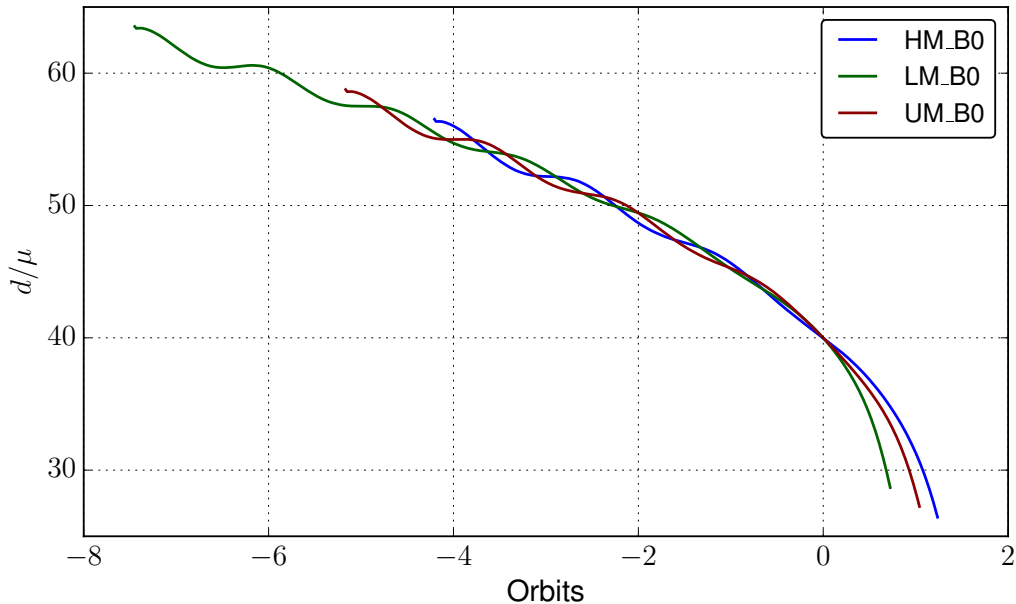

  \centering
  \includegraphics[width=0.9\columnwidth]{{{proper_sep}}}
  \caption{Proper separation between the centers of mass of the stars
    versus orbital phase.  For each model, the separation is given in
    units of the reduced gravitational mass $\mu$.  The orbital phases
    have been aligned to be zero at a separation of $40 \usk \mu$.}
  \label{fig:prop_sep}
\end{figure}

The HM, UM, and LM models complete 5, 6, and 8 orbits before merger,
respectively.  The inspiral is depicted in \fref{fig:prop_sep} in
terms of proper separation versus orbital phase.  We recall that for
point particles without spin and eccentricity, the separation scales 
with the reduced mass $\mu= M_g^1 M_g^2 / (M_g^1 + M_g^2)$, but also 
depends on the symmetric mass ratio $\nu=\mu/(M_g^1 + M_g^2)$.
For the unequal mass model, $\nu=0.2494$, which is very close to the 
value $\nu=0.25$ for equal masses. Ideally, the curves for all our 
models should only differ due to tidal effects. However, the 
differences during most of the inspiral are clearly dominated by the 
residual eccentricity of our initial data. Only during the last orbit, 
finite size effects become large enough to expose a trend: it seems 
that for the lighter models, the separation decreases more quickly 
with increasing orbital phase.

\begin{figure*}[t]
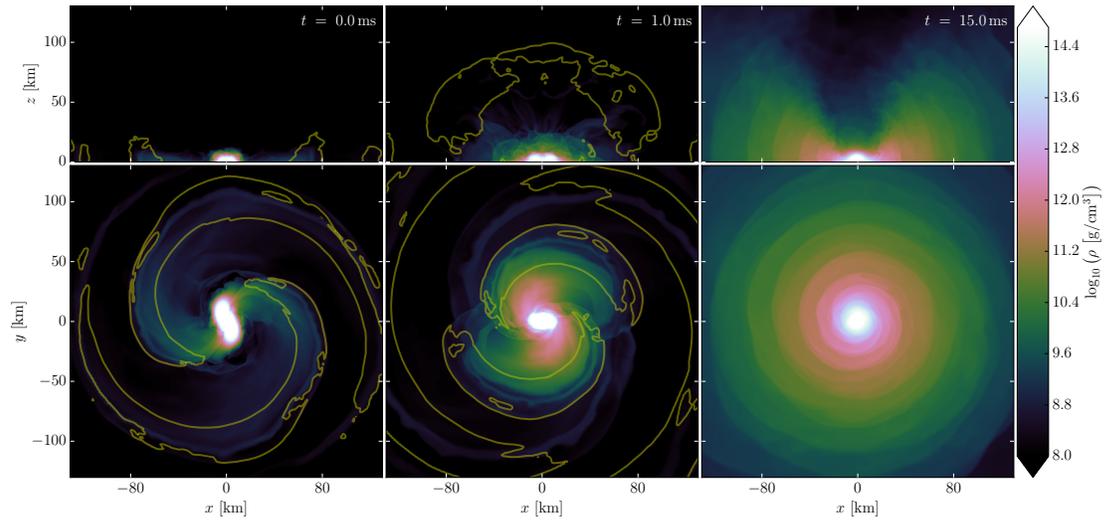

  \centering
  \includegraphics[width=0.95\linewidth]{{{cuts_snapsh_LM_B0}}}
  \caption{Snapshots of the evolution at $0,1$, and
    $15\usk\milli\second$ after the merger, for model \texttt{LM\_B0}.
    The top row show cuts in the $xz$-plane, the bottom row cuts in
    the orbital $xy$-plane.  The color corresponds to the logarithmic
    rest mass density. The contour lines mark the boundaries of
    regions where matter is unbound (according to the geodesic
    criterion).}
  \label{fig:snapshots_LM}
\end{figure*}

\begin{figure*}[t]
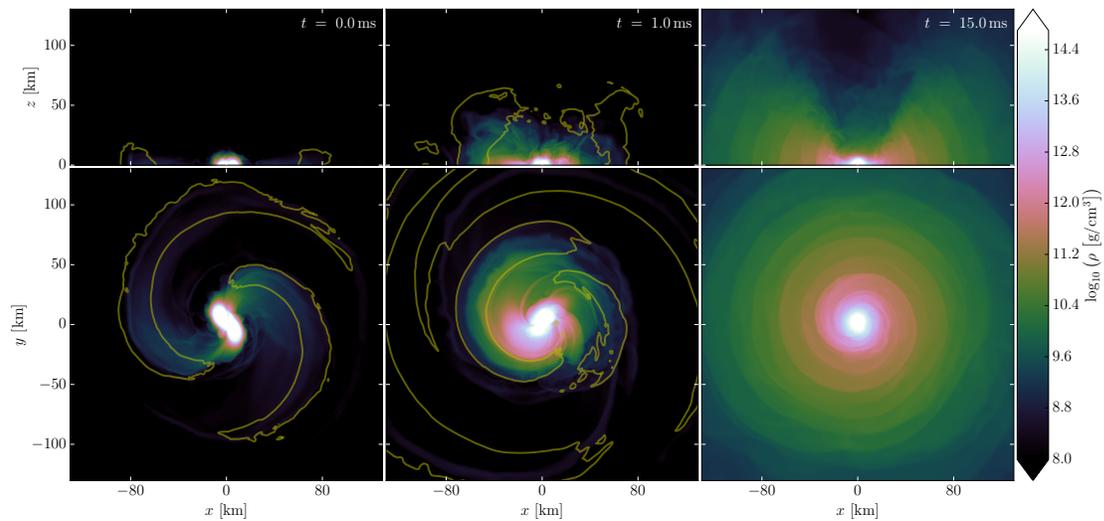

  \centering
  \includegraphics[width=0.95\linewidth]{{{cuts_snapsh_UM_B0}}}
  \caption{Like \fref{fig:snapshots_LM}, but for model
    \texttt{UM\_B0}. In the lower left panel, showing the time of
    merger, the lighter star is on the lower right side of the origin.}
  \label{fig:snapshots_UM}
\end{figure*}

\begin{figure*}[t]
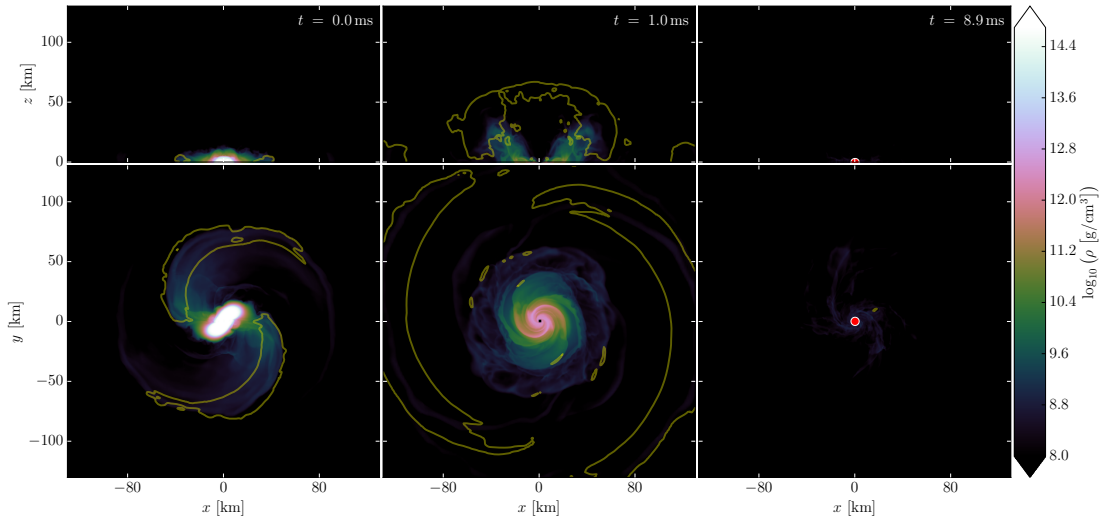

  \centering
  \includegraphics[width=0.95\linewidth]{{{cuts_snapsh_HM_B0}}}
  \caption{Like \fref{fig:snapshots_LM}, but showing model
    \texttt{HM\_B0}. The apparent horizon and its interior are drawn
    in white and red, respectively.}
  \label{fig:snapshots_HM}
\end{figure*}

The evolution starting at the merger is visualized in
figures~\ref{fig:snapshots_LM}, \ref{fig:snapshots_UM}, and
\ref{fig:snapshots_HM}.  All models tidally eject some matter shortly
before the merger. Not surprisingly, the total ejected mass for the
high mass model is negligible and the unequal mass model ejects more
than the low mass model. This will be discussed in \sref{sec:ejecta}.
The high mass model undergoes prompt collapse to a BH upon merger. The
BH mass and spin are given in \tref{tab:outcome}.  The low- and
unequal-mass models form supramassive remnants which are stable on the
timescale of our evolution, i.e. for more than $15\usk\milli\second$.

\subsection{Disk Structure}
\label{sec:disk}

In the following, we will describe the distribution of bound matter
outside the remnants.  Unbound matter will be discussed separately in
\sref{sec:ejecta}.  For the low- and unequal-mass models, the remnants
are surrounded by heavy and thick Keplerian disks.  As will be shown in
\sref{sec:rotprof}, the outer layers of the star already approach
Keplerian velocity. The transition between star and surrounding disk
is smooth (cf. \fref{fig:snapshots_LM}).

At the end of our simulation, the bound mass still contains a
significant fall-back component, i.e. matter that is moving along
highly eccentric trajectories and will eventually fall back onto the
remnant or the surrounding disk. In the equatorial plane, the
fall-back component is mainly composed of tidally ejected matter. We
found that its specific angular momentum is fairly constant.  In
detail, we compute the density-weighted $\phi$-average of specific
angular momentum in the orbital plane, $l(r)$.  For model
\texttt{LM\_B0}, the average of $l(r)$ taken over circumferential
radii $>50\usk\kilo\meter$ is $l_f = 7.0 \usk M_\odot$. We found a 
very similar value of $l_f = 7.2$ for model \texttt{UM\_B0}. For both,
the $L_1$-norm of the residual is below 7\% of $l_f$.   
The transition between disk and fall-back component is gradual. As a
ballpark figure, we note that the angular velocity 
profiles for Keplerian motion and for constant specific angular
momentum $l_f$ cross each other at a circumferential radius of
${\approx}30\usk\kilo\meter$ (the corresponding orbital period is
around $2\usk\milli\second$). Further out, the fluid flow becomes
gradually less stationary.

To quantify the mass distribution, we compute histograms of total and
unbound baryonic mass (excluding artificial atmosphere) with bins
corresponding to the coordinate radius.
From this, we can compute the total bound mass outside a given radius.
We also keep a histogram of proper volume, which allows us to define a
volumetric radius $r_v$ for the spheres of constant coordinate radius,
thus reducing the gauge ambiguities.
 
Although there is neither a clear distinction between disk and
fall-back component, nor between remnant and disk, we provide in
\tref{tab:outcome} the mass $M_d$ between $20 < r_v <
60\usk\kilo\meter$ as a tentative measure for the disk mass, and the
mass $M_f$ at $r> 60\usk\kilo\meter$ as a measure for the fall-back
component (note that the value for model \texttt{UM\_B0} is missing
simply because the simulation was performed before the introduction of those
measures).  The disk masses given in \tref{tab:outcome} are evaluated
$15\usk\milli\second$ after the merger. At this time, they are already
stationary. On the timescale of our simulations, we observe no
significant accretion onto the NS, and the expulsion of matter from
the NS due to its oscillations ceases around $10\usk\milli\second$.
Note that the remnants for the low mass models are in the mass range
of stable NSs and will survive for at least an accretion timescale
(but probably much longer).

The addition of the magnetic field apparently leads to a reduction of
the disk mass by ${\approx}31\%$ for the low-mass models (see
\tref{tab:outcome}). However, the disk masses directly after the merger
are almost identical. The differences only appear around
$5\usk\milli\second$ after the merger. At this point, the mass
ejection seems to be very sensitive to small changes, such as the
presence of a magnetic field.  The impact of the magnetic field on the
specific angular momentum of the fall-back matter is very weak: $l_f$
changes less than 1\% (2\%) for the low-mass (unequal mass) models.

In the high mass case, most material is swallowed immediately at
merger time when the BH is formed. The total mass remaining outside
the horizon at the end of the simulation is around $10^{-3}\usk
M_\odot$.

\begin{table}[t]
\caption{Outcome of the mergers.  $M_\mathrm{e}$ is our best estimate
  for the total ejected mass, and $v_\mathrm{esc}$ the average escape
  velocity (see text).  $f_\mathrm{pk}$ is the GW instantaneous
  frequency at merger time.  If a BH is formed, $M_\mathrm{BH}$ and
  $J_\mathrm{BH}$ are its mass and angular momentum, extracted at the
  end of the simulations.  $M_\mathrm{f}$ is the mass outside the
  apparent horizon.  For the models without BH, $F_\mathrm{c}$ and
  $F_\mathrm{m}$ denote the remnant's central and maximum rotation
  rates, computed $15\usk\milli\second$ after the merger.
  $M_\mathrm{d}$ and $M_\mathrm{f}$ are tentative measures for disk
  and fall-back masses, respectively (see text).  Finally,
  $f_\mathrm{pm}$ is the frequency of the largest peak in the post
  merger spectrum.}
\begin{indented}\item[]\begin{tabular}{@{}lllllll}\br 
Model&
\texttt{HMB0}&
\texttt{HMB13}&
\texttt{LMB0}&
\texttt{LMB13}&
\texttt{UMB0}&
\texttt{UMB13}
\\\mr
$M_\mathrm{BH}\,[M_\odot]$&
$2.79$&
$2.79$&
---&
---&
---&
---
\\
$J_\mathrm{BH}/M^2_\mathrm{BH}$&
$0.78$&
$0.78$&
---&
---&
---&
---
\\
$F_\mathrm{c}\, [\mathrm{kHz}]$&
---&
---&
$0.52$&
$0.49$&
$0.67$&
$0.66$
\\
$F_\mathrm{m}\, [\mathrm{kHz}]$&
---&
---&
$1.50$&
$1.54$&
$1.63$&
$1.60$
\\
$f_\mathrm{pk}\, [\mathrm{kHz}]$&
$2.18$&
$2.18$&
$2.02$&
$2.02$&
$2.08$&
$2.07$
\\
$f_\mathrm{pm}\, [\mathrm{kHz}]$&
---&
---&
$3.17$&
$3.14$&
$3.30$&
$3.26$
\\
$E_\mathrm{GW}\,[M_\odot]$&
$0.039$&
$0.039$&
$0.053$&
$0.056$&
$0.087$&
$0.080$
\\
$M_\mathrm{e}\, [M_\odot]$&
$<10^{-3}$&
$<10^{-3}$&
$0.002$&
$0.003$&
$0.010$&
$0.010$
\\
$v_\mathrm{esc}\, [c]$&
---&
---&
$0.13$&
$0.12$&
---&
$0.12$
\\
$M_\mathrm{d}\,[M_\odot]$&
---&
---&
$0.130$&
$0.091$&
---&
$0.119$
\\
$M_\mathrm{f}\,[M_\odot]$&
$0.001$&
$0.001$&
$0.085$&
$0.085$&
---&
$0.112$
\\\br
\end{tabular}\end{indented}
\label{tab:outcome}
\end{table}

\subsection{SMNS Rotation Profile}
\label{sec:rotprof}

In the following, we discuss the rotation profile of the remnant. In
particular, we are interested in understanding how strongly the
presence of the magnetic field affects the evolution of 
differential rotation.

\begin{figure}[t]
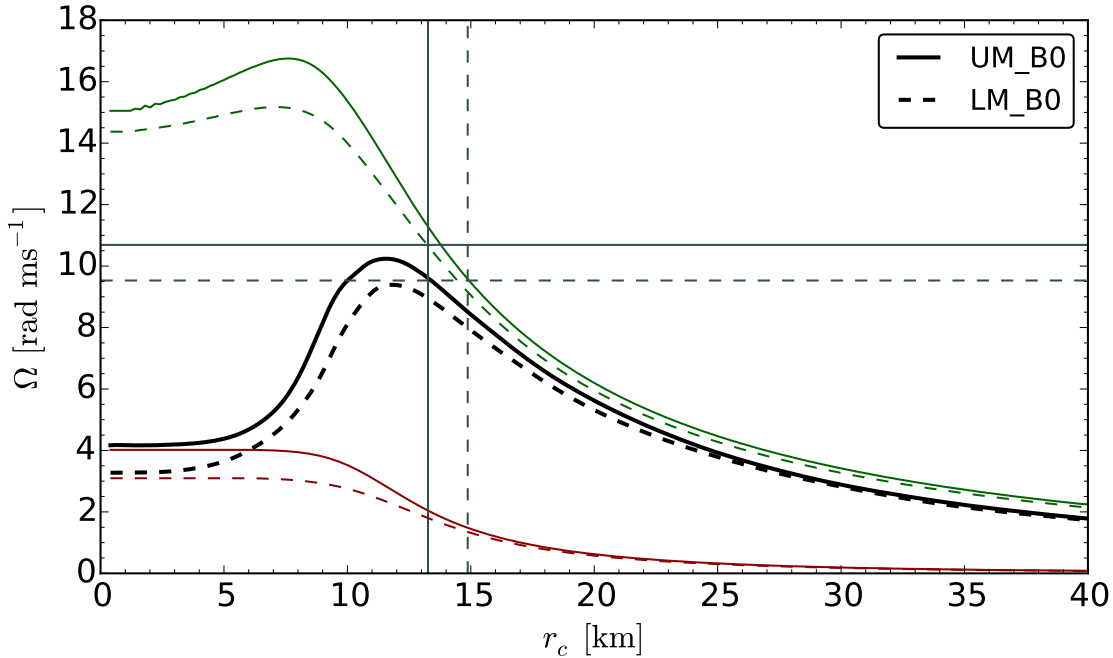

  \centering
  \includegraphics[width=0.99\columnwidth]{{{rotation_profiles}}}
  \caption{Rotation profiles after the remnant has settled down,
    averaged in time $14 \ldots 18\usk\milli\second$ after the merger.
    The thick black lines are the $\phi$-averaged rotation rates in
    the equatorial plane versus the circumferential radius. The red
    lines show the contribution of the frame dragging effect. The
    green lines shows the orbital angular velocity of a test particle
    in circular, co-rotating orbit. The grey horizontal line marks the
    pattern angular velocity of the $m=2$ component of the multipole
    decomposition of the mass density in the equatorial plane. The
    grey vertical line marks the radius where the density drops to 5\%
    of the central one.}
  \label{fig:rot_prof}
\end{figure}

To extract the rotation profile in a well defined way, we use the
coordinate system introduced in \cite{Kastaun:2015:064027}, which is
defined by a prescription independent of the spatial gauge used in the
numerical evolution. However, the prescription still requires the
choice of the origin. If the spacetime is approximately axisymmetric
around the origin, the resulting coordinate system will reflect that fact,
with $\phi$ coordinate lines approximating Killing vectors. For our
purpose, the origin should obviously be located at the center of
rotation of the remnants.  For equal mass models, the symmetry axis
defines an origin in a gauge independent way.  For the unequal mass
models, we need to compensate for the residual movement of the
remnant.  Therefore, we use the center of mass (CMS) computed in simulation
coordinates as the origin.  Note that the CMS, and hence our choice of
origin, is not a gauge invariant definition, and in contrast to the
Newtonian case oscillations can weakly influence the CMS position.  In
order to prevent any feedback of oscillations, we use the running
average in time of the CMS position, with a smoothing length of
$4\usk\milli\second$.

From our simulations, we obtain qualitatively similar rotation profiles 
as presented in \cite{Kastaun:2015:064027} for different models.
In particular, this applies also for the unequal-mass model, while 
\cite{Kastaun:2015:064027} was restricted to equal-mass models. 
After the merger, the remnant
settles down within a few $\milli\second$ to a state with a slowly
rotating core, a maximum in the outer layers, and at larger radii a
slow falloff towards the Keplerian velocity. This is depicted in
\fref{fig:rot_prof}.  We note that directly after the merger, the
maximum rotation rate is at the center. However, this configuration
seems to be unstable and quickly changes towards the final profile.
The nature of this transition is still under investigation, but it
is likely connected also to the Kelvin-Helmholtz (KH) instability.
It is worth noting that the pattern angular velocity of the $m=2$ 
moment of the density is very similar to the maximum of the rotation 
rate. This yet unexplained correlation was already observed in 
\cite{Kastaun:2015:064027}.

\begin{figure}
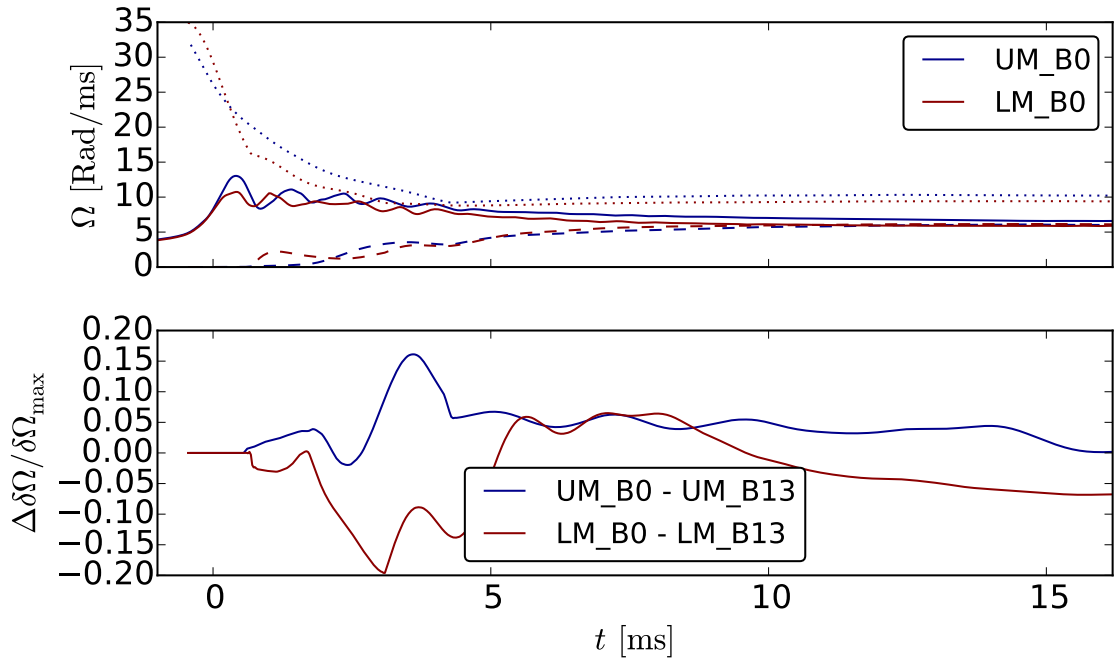

  \centering
  \includegraphics[width=0.99\columnwidth]{{{hmns_diff_rot}}}
  \caption{Evolution of rotation profile properties.  The solid line
    shows the average rotation rate (see main text).  The dotted line
    shows the central rotation rate, starting at the time the two
    stars touch (more precisely, when the central density reaches 5\% of the
    maximum one).  The dashed line shows the differential rotation in
    terms of the difference between maximum and central rotation rate.
    To highlight the overall evolution of central and differential
    rotation, we suppressed the influence of short term oscillations
    by smoothing in time via convolution with a Gaussian of width $0.5
    \usk\milli\second$.  }
  \label{fig:diff_rot}
\end{figure}

The rotation profiles for magnetized and non-magnetized models differ
only slightly. The profile for the magnetized model \texttt{UM\_B13}
differs from the non-magnetized one depicted in \fref{fig:rot_prof} by
less than 3.2\%, with an average difference of 1.3\%.  In particular,
differential rotation is not reduced significantly.
\Fref{fig:diff_rot} depicts the evolution of the maximum rotation rate
as well as the difference in rotation rate between the center and the
maximum.  The differential rotation defined above is zero initially
because the central rotation rate is also the maximum one at the
merger. This changes rapidly, within ${\approx}5\usk\milli\second$.
Subsequently, the differential rotation keeps increasing slowly
towards the final value.  Note that the influence of the magnetic field
seems to be stronger during the rapid rearrangement phase. This fits
the conjecture that it is some sort of instability, and as such the
exact time evolution would be sensible to the seed perturbations, but
not the final state.

\Fref{fig:diff_rot} also contains the density weighted average of the
rotation rate in the equatorial plane. During the first
${\approx}5\usk\milli\second$ after the merger, the rotation rate
oscillates, with decaying amplitude. This is caused by the
quasi-radial oscillations excited at merger, which change the
moment of inertia periodically. Since rotation and differential
rotation approach stationarity, we report the final values for all
models in \tref{tab:outcome}.



\begin{figure}[t]
  \centering
  \includegraphics[width=0.99\columnwidth]{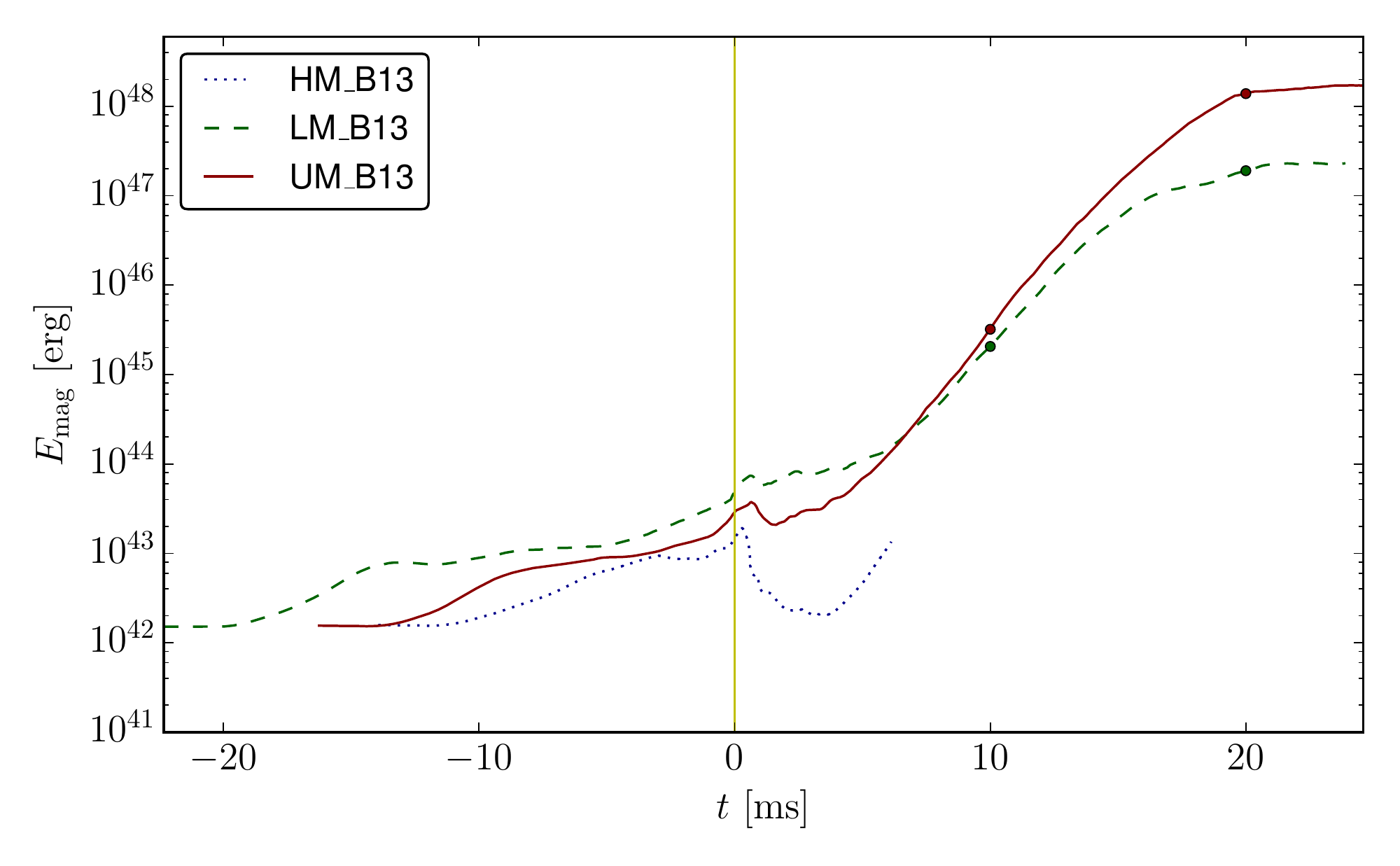}
  \caption{Evolution of the total magnetic energy for the three
    magnetized models: high-mass (blue dotted), low-mass (green
    dashed), and unequal-mass (solid red). The vertical line marks the time
    of merger $t=0$. Circle markers indicate the times of the 
    snapshots shown in \fref{fig:2D_B}.}
  \label{fig:Emag}
\end{figure}

\subsection{Magnetic Field Evolution}
\label{sec:mag}

\noindent We now turn our attention to the evolution of magnetic
fields. 
\Fref{fig:Emag} depicts the evolution of total magnetic energy. 
As one can see, there is a moderate amplification already during the 
inspiral. There are several possible effects that might contribute to 
the evolution of the field. First, the chosen field configuration is 
known to be unstable and might re-arrange itself, which is however 
unlikely to amplify the field. A second possible cause could be fluid 
flows induced by tidal forces or GR effects, which is however purely 
speculative. A more likely cause is the imperfection of the initial 
data. The error due to the quasi-circular approximation might lead to 
some vortex-like, churning movements on top of more visible effects 
such as residual eccentricity and stellar oscillations. Numerical 
errors during the evolution can be ruled out as cause of the 
amplification, since we observe slightly more amplification during 
inspiral with better resolution, not less. In summary, we cannot tell 
if the amplification during inspiral is a generic feature or an 
artifact of our setup. In any case, we are not overly concerned about 
the changes during the inspiral since the actual field structure of BNS 
is completely unknown anyway and our setup is intended only as generic 
example of a magnetized merger. Moreover, we are confident that those 
changes do not influence the qualitative results in the post-merger 
phase which will be discussed in the following.

\begin{figure*}[t]
  \centering
  \includegraphics[width=0.99\columnwidth]{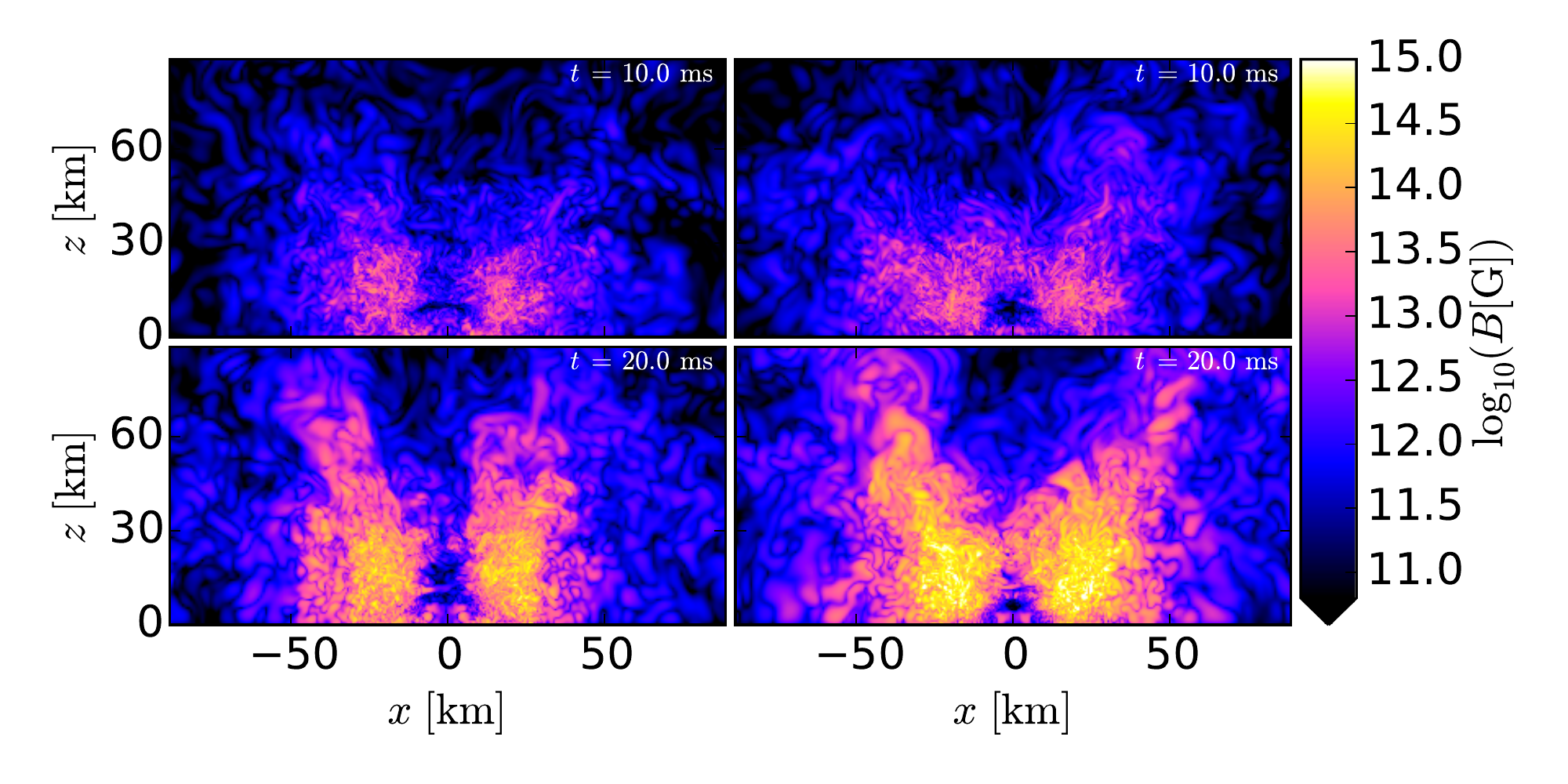}
  \caption{Meridional view of the magnetic field strength at selected
    times ($10$ and $20\usk\milli\second$ after merger) for the 
    low-mass model (left) and the unequal-mass model (right). }
  \label{fig:2D_B}
\end{figure*}

For the high-mass model, most of the matter and the associated magnetic energy 
is swallowed immediately at merger when the BH forms. The increase in magnetic 
energy a few $\milli\second$ after BH formation seems to be caused by the
remaining low-density matter that is falling back onto the BH. 

The magnetic energy of the low- and unequal-mass models undergoes an 
exponential growth phase after merger that lasts 
${\sim}10$--$15\usk\milli\second$, with an e-folding time of
$\tau\approx1.3\usk\milli\second$. Around $20\usk\milli\second$ after 
merger, the fields have reached a saturation. At this point, the magnetic energy 
in the unequal mass model has increased by almost five orders of magnitude 
compared to the energy at merger time. The main difference between the two
cases is that the saturation amplitude we observe for the low-mass model 
is around one order of magnitude smaller. In addition, the growth rate
is slightly lower. For the unequal-mass model, we performed a convergence
test (see \ref{sec:app}), which showed that the magnetic field amplification 
is still under-resolved. For a higher resolution, we find a saturation amplitude 
that is larger by one order of magnitude. 

The spatial distribution of the magnetic field strength in the meridional
plane is depicted in \Fref{fig:2D_B}. It shows two snapshots, one at
$10\usk\milli\second$, in the middle of the exponential growth phase, and
one at $20\usk\milli\second$, when the fields have reached saturation 
amplitude.
Interestingly, the field near the rotation axis is much stronger for the 
unequal-mass model than for the low-mass model. Also in relation to the 
field in the torus, the region near the axis is magnetized more strongly
for the unequal mass case. The field is however unordered and not suitable
to produce a jet. Furthermore, the level of baryon pollution along the
axis differs. The rest-mass density along
the axis can be up to one order of magnitude higher in the unequal-mass
case (cf. figures \ref{fig:snapshots_LM} and \ref{fig:snapshots_UM}).  

As shown in \fref{fig:2D_B}, the maximum magnetic field strengths occur
${\sim}20\usk\kilo\meter$ away from the orbital axis, a region 
with significant negative gradient of the angular velocity (cf. 
\fref{fig:rot_prof}). Thus, magnetic winding should be at least partially 
responsible for the amplification. However, we found that toroidal and 
poloidal fields reach comparable strength in the torus. This might be
explained by an unordered component of the fluid flow in the torus on top
of the overall differential rotation, providing an effective redistribution 
of magnetic energy in the two components. 
Despite the amplification of magnetic fields, we find that the 
magnetic-to-fluid pressure ratio stays below $10^{-2}$ everywhere 
up to the end of the simulation, indicating that magnetic fields 
remain always dynamically subdominant in our simulations (see 
\ref{sec:app_mag} and \fref{fig:MHD_ratio}).

\begin{figure*}[t]
  \centering \includegraphics[width=0.99\linewidth]{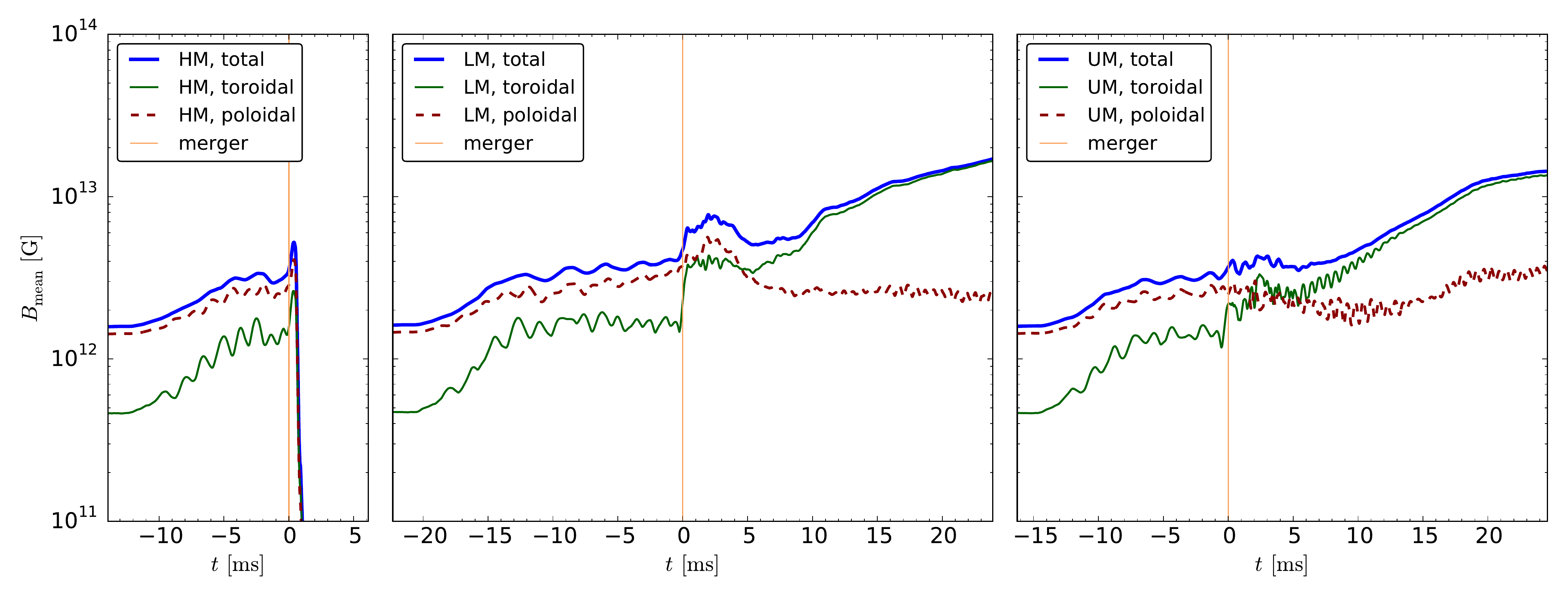}
  \caption{Evolution of the density-weighted average of the magnetic
    field strength (see text) for the three magnetized models:
    high-mass (left), low-mass (middle), and unequal-mass (right). In
    addition to the total strength (thick blue line), toroidal and
    poloidal field components are plotted (green and dashed red lines,
    respectively). The vertical line marks the time of merger $t=0$.
  }
  \label{fig:Bmean}
\end{figure*}

As a measure for the evolution of the magnetic field inside the remnant,
we use the density-weighted average of the magnetic field strength, 
$B_\mathrm{mean}\equiv \int \rho B dV \, /\, \int \rho dV$. In the 
same way, we also compute the averages of the poloidal and toroidal
components. The evolution of the averages is shown in \fref{fig:Bmean} 
for the three magnetized cases (HM, LM, UM).
As discussed earlier, there is already some amplification during 
the inspiral, which is most likely a consequence of our initial field 
choice. From \fref{fig:Bmean}, we find that the field at merger is 
still mostly poloidal for all three models. Since the high-mass model
promptly collapses into a BH, we only discuss the other two cases in 
the following.

After merger, the average magnetic field undergoes irregular variations
for 3--5~$\milli\second$ and then it starts to increase continuously, 
although in the UM case it starts to saturate near the end of the 
simulation. The 
overall amplification in the remnant is much smaller than in the disk. At
$t = 20\usk\milli\second$, the average field strength is larger by a 
factor ${\approx}3$ compared to the time of merger. 
In the LM case, we observe a steep temporary increase when the two stars 
touch. This might be due to the KH instability, which 
we expect to develop in the shear layer between the two stars, producing 
vortices in the orbital plane and amplifying the toroidal magnetic 
field component. However, it is difficult to draw conclusions from 
$B_\mathrm{mean}$ directly after merger because the density also 
undergoes rapid changes during this period, thus changing the relative 
weight of different parts of the field.
In the UM case, the evolution of the average magnetic field is  
smoother. Nevertheless, when repeating the simulation with a higher resolution, 
we found a temporary post-merger amplification by a factor $\sim2$ (see 
\fref{fig:B_conv}), more similar to the low-mass case in \fref{fig:Bmean}.
Recent work confirmed that a resolution
of few tens of meters or better (much higher than the one employed here) 
is required in order to properly resolve the small scales at which the KH
instability is most effective
\cite{Kiuchi:2015:1509.09205}. Our results thus under-estimate the 
amplification due to the KH instability.

Magnetic winding on the other hand should be well resolved. Since
the merger remnants are differentially rotating, as shown in 
\fref{fig:rot_prof}, we expect an amplification of the toroidal 
field inside the SMNSs. Indeed, by comparing the density-weighted 
averages of poloidal and toroidal field shown in \fref{fig:Bmean},
we find that mainly the toroidal field is amplified inside the SMNS.
Note that although the magnetic winding occurs at the expense of 
differential rotation, the rotation profiles are almost unaffected 
at the given field strength, as discussed in \sref{sec:rotprof}.  

In theory, the magneto-rotational instability (MRI)
\cite{Balbus:1991} is an additional powerful amplification mechanism
that could act inside the bulk of the SMNS as well as in the accretion
torus (e.g.~\cite{Siegel:2013:121302,Kiuchi:2014:41502}), although our
present resolution is insufficient to properly resolve the wavelength
of the fastest growing MRI mode.

Our LM and UM simulations show no indication in favor of the
formation of a relativistic jet or any kind of outflow along the
orbital axis, suggesting that these systems could not be responsible
for SGRBs. This is in agreement with the general expectation that,
while the formation of a BH-torus system within ${\sim}100\usk\milli\second$ after
merger could provide the necessary conditions to launch a relativistic
jet, the formation of a long-lived remnant NS can hardly act as a SGRB
central engine, mostly due to the strong baryon pollution along the
orbital axis 
\cite{Dessart:2009:1681,Hotokezaka:2013:24001,Siegel:2014:6,Siegel:2015:procs}.

As a side note, an alternative possibility is the formation of a
relativistic jet at a later stage, when the SMNS eventually
collapses to a BH, as envisaged in the ``time-reversal" scenario
\cite{Ciolfi:2015:36}.
Before a SMNS can collapse, it has to loose a significant fraction of its 
angular momentum. Apart from the angular momentum carried away via GWs
after merger, a good fraction of the rotational energy can be emitted via
spin-down radiation, powering a potentially strong and long-lasting
EM signal. This signal constitutes a promising
counterpart to the gravitational wave emission from BNS mergers (see
\cite{Siegel:2016a,Siegel:2016b} and references therein).

We stress again that due to the under-resolved KH and MRI mechanisms,
the field amplification we observe can only act as a lower
limit. In particular, KH instability is resolved, but only for
vortices with length scale around at least $5$ times our resolution
(see
also \cite{Baiotti:2008:84033}, \cite{Giacomazzo2011PhRvD..83d4014G}). Smaller
vortices would need higher resolution to contribute to the magnetic
field amplification. Instead, a resolution $\approx 100$ times higher
would be necessary in order to start resolving the MRI and see its
effects on magnetic field amplification. In particular, we cannot
rule out jet formation completely, although we consider it unlikely
for our models. In addition, the prescription for the initial field
might lead to differences. For example, our magnetic fields are
initially confined inside the stars, while the presence of a global
poloidal field extending outside the stars prior to merger might favor
the formation of a magnetic funnel \cite{Paschalidis:2015:14}.



\begin{figure*}[t]
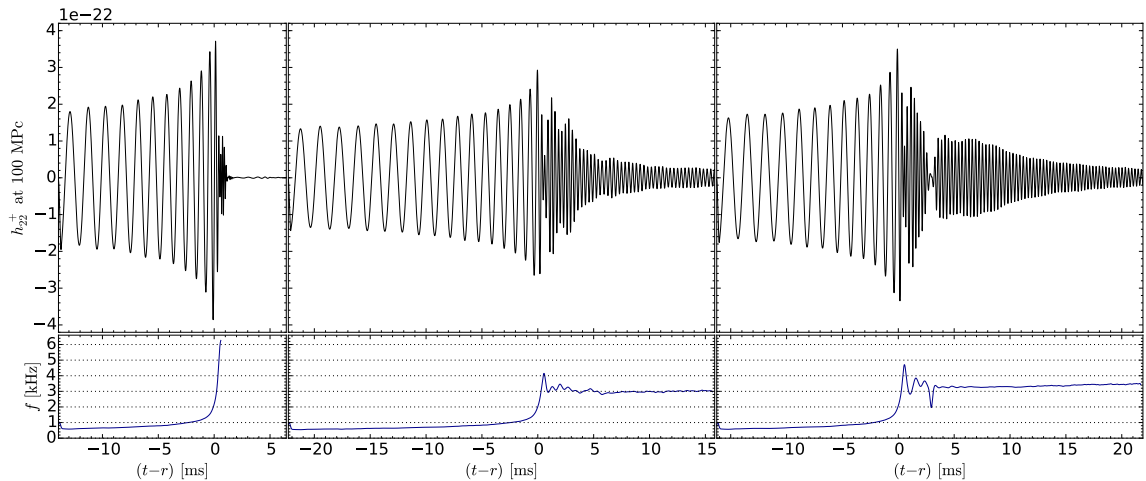

  \centering
  \includegraphics[width=0.99\linewidth]{{{hp_multisim_l2m2}}}
  \caption{Gravitational wave signal for models \texttt{HM\_B0}, 
    \texttt{LM\_B0}, and \texttt{UM\_B0} (from left to right).  
    The top panels show the strain at nominal distance of
    $100\usk\mega\parsec$.  The lower panels show the instantaneous
    frequency.}
  \label{fig:gw_strain}
\end{figure*}

\subsection{Gravitational Wave Signal}
\label{sec:gw}

We extract the GW signal for all runs at a fixed radius of $738
\usk\kilo\meter$, using the Moncrief formalism (we also use the
Weyl scalar $\Psi_4$ as a crosscheck).  We perform no extrapolation to
infinity, since the precision is likely limited by the accuracy of the
hydrodynamic evolution.
Throughout this section, the strain will be given in terms of the
coefficients $h_{lm}$ of the expansion in spin weighted spherical
harmonics (see \cite{Thorne:1980:299}), where $h_{lm} = h^+_{lm} + i
h^\times_{lm}$.  The actual strain at a given viewing angle to the
rotation axis is obtained by multiplication with the spin weighted
spherical harmonic $|{}_{-2}Y_{22}(\theta,\phi)|$.
  
The $l=m=2$ component of the strain for the non-magnetized models is
shown in \fref{fig:gw_strain}.  The high mass model collapses to a BH
within $1\usk\milli\second$ after merger, hence the GW signal consists
mainly of the inspiral part plus a short BH ringdown.  The low- and
unequal-mass models do not collapse and exhibit a strong post-merger
signal.

The other components of the strain are smaller than $7\%$ of the
dominant $l=m=2$ component in all cases.  For the high mass model, the
second largest contribution comes from the $l=m=4$ component, with a
relative amplitude of $5.3\%$. The other components are smaller by a
factor at least $5$ apart from $l=3, m=2$ which is only smaller by a
factor $2$.  By comparing the spectra, we found that the $l=4$
component is just an overtone of the $l=2$ component, i.e. the peak
frequency matches.  For the low-mass model, the $l=m=4$ component has
a relative amplitude of $4.1\%$. More interesting, we observe a
growing $l=m=3$ component after the merger. Its maximum amplitude is
$1.2$ of the $l=m=4$ maximum, but it is reached $29.2\milli\second$
after the merger.

\begin{figure}[t]
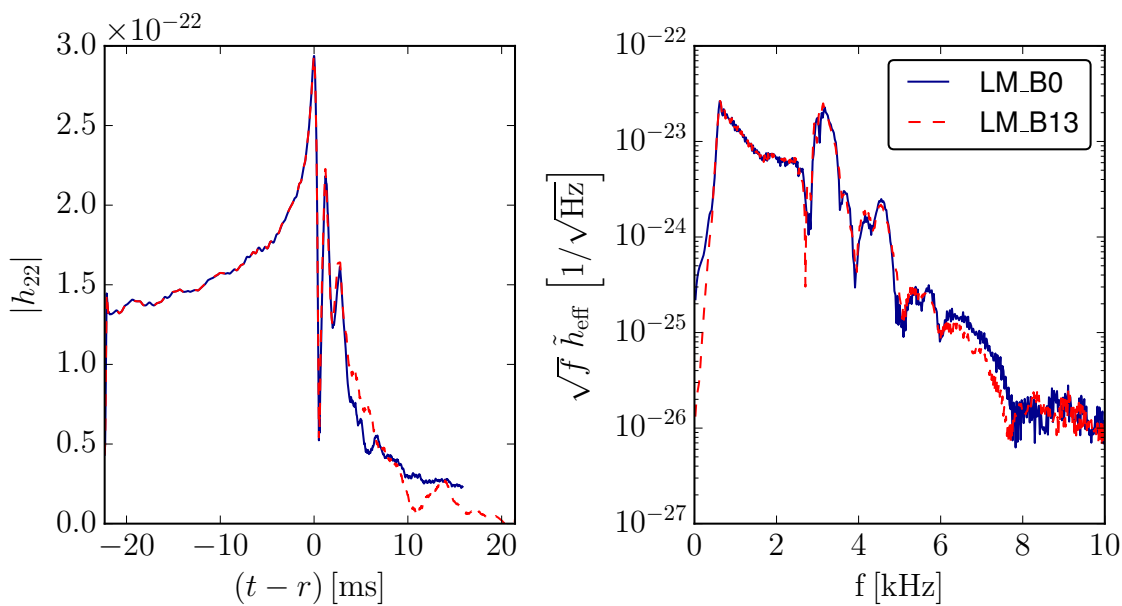

  \centering
  \includegraphics[width=0.99\columnwidth]{{{LM_B0vB13}}}
  \caption{Influence of magnetic field on gravitational wave strain
    amplitude (left panel) and spectra (right panel) for the low-mass
    models. The strain is given at distance of $100
    \usk\mega\parsec$.}
  \label{fig:gw_LM_Bcomp}
\end{figure}

The differences in the strain between magnetized and non-magnetized
models during the inspiral are negligible.  This is not surprising
since the energy in the magnetic field in the orbiting NSs is below
$10^{-11}$ of a single star's gravitational binding energy.  In the
post-merger phase however, we observe some differences. For the
unequal-mass model, the amplitude differs by up to 8\% of the maximum
one.  The phase difference steadily grows up to $6\usk\mathrm{Rad}$ at
the end of the simulation.  The knot visible in the amplitude is
present in both cases.  For the low-mass model, the strain amplitude
for the magnetized model exhibits a knot
${\approx}10\usk\milli\second$ after the merger, which is not present
in the non-magnetized case. This is shown in \fref{fig:gw_LM_Bcomp}.
At this time the difference in amplitudes is maximal, around 10\% of
the maximum amplitude (at merger time). Further, around the knot the
phase shift quickly grows to around $3.5\usk\mathrm{Rad}$.  The
influence on the spectrum is however small, as shown in
\fref{fig:gw_LM_Bcomp}.

\begin{figure}[t]
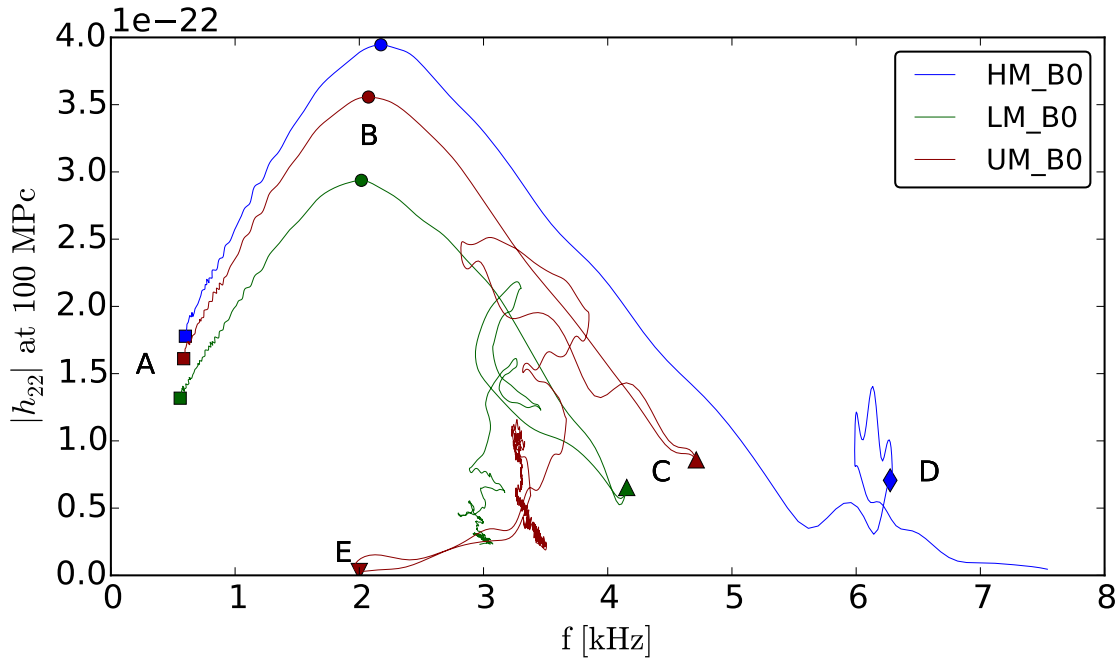

  \centering
  \includegraphics[width=0.99\columnwidth]{{{h_to_f_multisim_l2m2}}}
  \caption{Gravitational wave amplitude $|h_{22}|$,
  versus the instantaneous frequency of the signal.
  The markers correspond to the following events: 
  (\textbf{A}) Inspiral,
  (\textbf{B}) Merger,
  (\textbf{C}) Maximum frequency (not for prompt collapse),
  (\textbf{D}) Apparent horizon formation,
  (\textbf{E}) Knot in the GW signal for the unequal-mass model.} 
  \label{fig:gw_amp_freq}
\end{figure}

To track the time evolution of the dominant GW component, we compute
its instantaneous frequency from the phase velocity of the complex
strain $h_{22}$. To prevent amplifying high-frequency noise we use a
Gaussian kernel derivative with smoothing length $0.1\usk\milli\second$.  
The instantaneous frequency time evolution is shown in \fref{fig:gw_strain}.
In \fref{fig:gw_amp_freq}, we plot this frequency versus the strain 
amplitude. The evolution
can be divided into several phases. During the inspiral (starting at
points labeled A), both frequency and amplitude increase, the latter
reaching a maximum when the stars merge.  While the merged object
becomes more compact, and hence rotates faster, the frequency keeps
increasing.  The amplitude on the other hand decreases again. This can
be explained by the fact that the remnant becomes both smaller and
more axisymmetric, hence reducing the quadrupole moment.  For the high
mass model, the system then undergoes collapse (point D marks the
formation of the apparent horizon), during which the amplitude raises
by a factor of 3 for a 
short time.  The other models exhibit a bounce (points C). While the
remnant expands again, both frequency decreases while the amplitude
increases up to around $70\%$ of the maximum amplitude.  The low-mass
model undergoes a second, smaller bounce. Afterwards, the oscillation
frequency remains stable, while the amplitude decays slowly.  The
unequal-mass model shows a curious knot ${\approx}3\usk \milli\second$
after the merger (point E; note the instantaneous frequency at this
point is meaningless because the amplitude of the dominant mode is
close to zero, and the phase of the strain is hence determined by
the other components and numerical errors).
Around $5\usk\milli\second$ after merger, however, the signal is  
that of one mode with decaying amplitude and slowly increasing frequency. 
We caution that the error for the GW amplitude after merger is unknown,
since we could not demonstrate numerical convergence for the post-merger 
GW amplitude (see \ref{sec:app}).
We also note that the signal is missing the long-term evolution (not
covered by the simulations), during 
which rotation rate, compactness, and frequency can still undergo slow, 
but significant drift.

\begin{figure}[t]
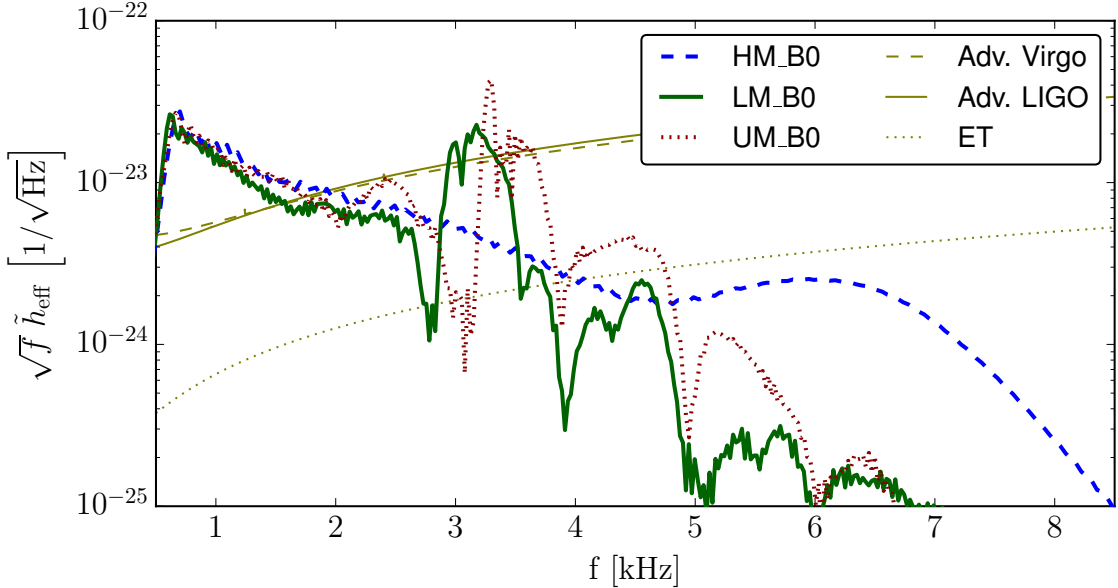

  \centering
  \includegraphics[width=0.99\columnwidth]{{{gw_spec_multisim_l2m2}}}
  \caption{Gravitational wave spectra (thick lines) for the three
    models in comparison to the sensitivity curves of GW detectors
    (thin lines). The strain is given at distance of 
    $100\usk\mega\parsec$.}
  \label{fig:gw_spec}
\end{figure}

We now discuss the power spectra of the GW signal, given by
$h_\mathrm{eff}(f) = \sqrt{\tilde{h}^2_+(f) + \tilde{h}^2_\times(f)}$,
where $\tilde{h}_+, \tilde{h}_\times$ are the Fourier transforms of
coefficients $h^+_{22}(t), h^\times_{22}(t)$.  The spectra for all
models are shown in \fref{fig:gw_spec}, in comparison to the
sensitivity curves of GW detectors.  At a distance of 100 Mpc, the 
inspiral phase of all our models will be visible with both advanced 
LIGO and Virgo, while the post-merger part of the spectrum for the 
low- and unequal-mass models will be barely visible.   
Note that our signal does not include the long term
evolution of the remnant.  Although the amplitude at the end of our
simulation is quite low, a longer integration time might enhance its
detectability (e.g. \cite{DallOsso:2015,Gualtieri:2011}). This depends
on the damping at late times and the 
stability of the frequency.  The high-frequency side-peak of the
unequal mass models will be barely visible with the Einstein
telescope, while the one produced by the low-mass models is too
faint. The frequency of the largest post-merger peak of the spectra
is given in \tref{tab:outcome} for each model, together with the
instantaneous frequency at merger time.

\subsection{Matter Ejection}
\label{sec:ejecta}

Computing the amount of ejected matter from numerical simulations is
not straightforward for several reasons.  First, it is very expensive
to run the simulation long enough to let all ejected matter reach very
large radii. We are therefore forced to apply approximate criteria to
determine if a fluid element will eventually escape. There are two
such criteria in use, both assuming a stationary spacetime. One is
based on the assumption of geodesic motion, which leads to the
condition $u_0 < -1$, where $u_\mu$ is the fluid 4-velocity and 
we assume that the lapse is normalized to unity at infinity. The other
is derived from the relativistic Bernoulli condition for stationary
fluid flows, expressed by a similar condition $h u_0 < -1$
(note $h$ is only defined up to a constant factor, here we use 
the convention that $h=1$ for cold, infinitely diluted matter).
The relativistic enthalpy $h$ is larger than one, hence the Bernoulli
condition always predicts a larger amount of unbound matter. In
particular, the estimates increase with temperature.

\begin{figure}[t]
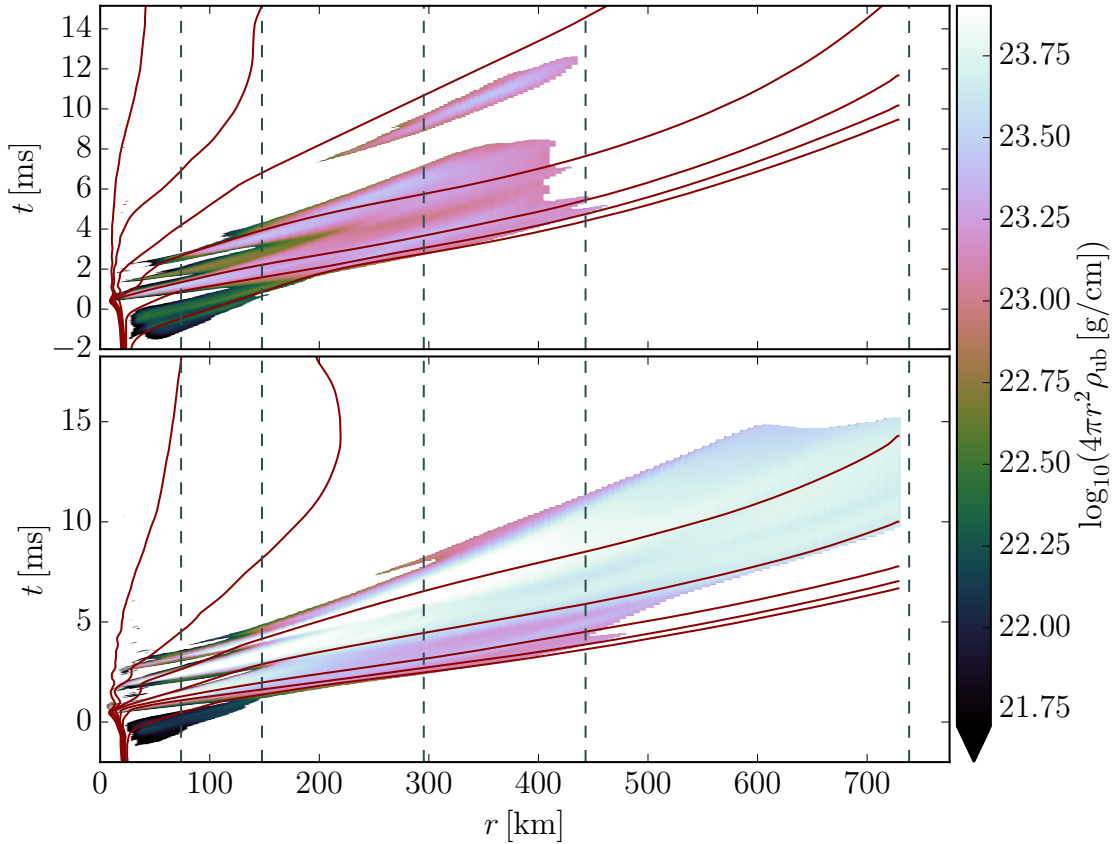

  \centering
  \includegraphics[width=0.99\columnwidth]{{{ejecta_rc_t}}}
  \caption{Distribution of matter versus radius and time, for models
    \texttt{LM\_B13} (top panel) and \texttt{UM\_B13} (bottom panel).
    The color corresponds to the amount of unbound matter in spherical
    shells per radius. Regions where the average density of unbound
    matter falls below the artificial atmosphere density are drawn in
    white.  The total mass (excluding artificial atmosphere) outside a
    given radius is depicted with contour lines corresponding to
    $10^{-n_i} \usk M_\odot$, where $n_i = 1, 1.5, \ldots, 4$, from
    left to right.  Note the contours in the lower panel become
    increasingly inaccurate after ${\approx}10\usk\milli\second$ due
    to matter reaching the boundary.  The vertical lines mark the
    position of the spherical surfaces we use to compute mass fluxes.}
  \label{fig:ejecta_rc_t}
\end{figure}

Since the volume integral of the unbound mass according to both
prescriptions is used in the literature, we made a comparison.  The
largest difference was found for model \texttt{LM\_B0}, where the
Bernoulli criterion predicts around twice as much ejected mass.  This
can be attributed to the inclusion of shock-heated material.  We
consider the geodesic assumption more realistic for our simulations
since the ejected matter is launched in expanding shells which do not
resemble a stationary fluid flow at all. It seems implausible that the
thermal energy of a thin shell can be used to accelerate it outwards
as it would be the case for fluid elements in a steady flow.
In the rest of this section, we use the geodesic criterion.
Of course, in the highly dynamic region near the remnant it is invalid
as well.

A second problem is caused by the decreasing density of the outflowing
matter, which eventually becomes less than the density of the artificial
atmosphere used in our numerical scheme, causing loss of unbound
matter at large radii. Fortunately, there is some freedom regarding
the method for integrating the local measures to get the total amount
of ejected matter, which we will exploit to minimize the influence of
the above problems.

To map out the evolution of the radial matter distribution we compute
at each time histograms of the amount of bound and unbound matter as
well as proper volume, all versus coordinate radius.  From this, we
compute the average density of unbound matter at each radius, as well
as the total amount of matter outside each radius.  This is shown for
two representative models in \fref{fig:ejecta_rc_t}.  As one can see,
the average unbound matter density drops below the artificial
atmosphere density before reaching $500\usk\kilo\meter$ for model
\texttt{LM\_B13}.  At this point, the measure becomes increasingly
meaningless (although that also depends on the degree of spherical
symmetry, with concentrated lumps of matter surviving longer).  For
model \texttt{UM\_B13} on the other hand, part of the ejected matter
escapes through the outer boundary.

\begin{figure}[t]
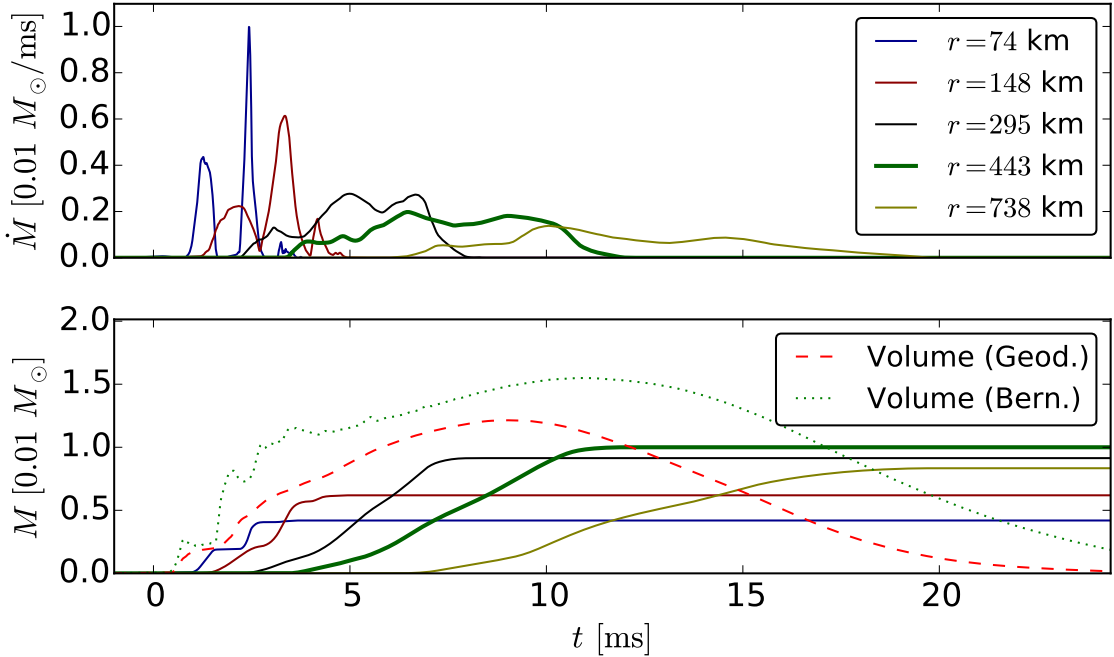

  \centering
  \includegraphics[width=0.99\columnwidth]{{{ejecta_measures}}}
  \caption{Comparison of different estimates for the unbound matter
    for run \texttt{UM\_B0}.  The top panel shows fluxes of unbound
    matter, according to the geodesic criterion, through spherical
    surfaces at various radii.  The bottom panel shows the time
    integrated fluxes from the top panel, and the volume integral over
    the computational domain for both geodesic and Bernoulli
    condition. The thick line is the one used for our best estimate.}
  \label{fig:unbound_measures}
\end{figure}

\Fref{fig:ejecta_rc_t} also reveals that the matter is ejected in
several waves, for both models. Using an animation, we identified the
nature of each wave.  For the unequal mass model, we find that the
first wave is tidally ejected shortly before the stars touch. This
wave has two arms, the one from the lighter star being slightly
larger. It is relatively slow, contains very little matter, and is
propagating in the orbital plane.  The second wave is a combination of
tidal ejection, expansion of the dense ejected matter, and shock
waves. It is emitted in a more isotropic fashion and contains more
matter. It is also faster, sweeping up the first wave.  The third wave
seems to be caused by shocks originating from the oscillating remnant.
The situation for the low mass model \texttt{LM\_B13} is qualitatively
the same, apart from the asymmetry in the unequal mass case.
 
As indicated by \fref{fig:ejecta_rc_t}, it can be difficult to find a
time where all ejected matter is neither close to the source nor
dissolving into the atmosphere. In this case, using volume integrals
to estimate the ejected mass is problematic.  A better option is the
use of the unbound matter flux through spherical surfaces at a
suitable radius, integrated in time.  Since the best choice is unknown
before the simulation, we monitor the fluxes through several surfaces,
with radii $r=74,148,295,443$, and $738 \usk\kilo\meter$ (cf.
\fref{fig:ejecta_rc_t}).  We then use the surface yielding the maximum
ejected mass for our best estimate.  For the unequal mass models the
corresponding radius is $r=443\usk\kilo\meter$, for the other models
$r=295\usk\kilo\meter$.  Additionally, we employ histograms like
\fref{fig:ejecta_rc_t} to ascertain that the ejected matter at those
radii are not too diluted.

Our best estimates are given in \tref{tab:outcome}, and a comparison
of the different measures is shown in \fref{fig:unbound_measures}.
Based on figures \ref{fig:ejecta_rc_t} and \ref{fig:unbound_measures},
we estimate the uncertainties for the ejected mass due to the
extraction method to be around $30\%$. In addition, we expect an error
due to the finite resolution around 50\%, as will be detailed in
\ref{sec:app}.  Not surprisingly, the high mass model shows no
significant matter ejection, and we only provide an upper limit. As
expected, the unequal mass model, although heavier, clearly ejects
more matter than the low mass model.

To estimate the escape velocity, we compute an average escape Lorentz
factor defined as $W_\infty = -\left( \int u_0 \rho_\mathrm{u}
\mathrm{d}V\right) / \left(\int \rho_\mathrm{u} \mathrm{d}V \right)$,
where $\rho_\mathrm{u}$ is the density of unbound matter and
$\mathrm{d}V$ is the proper volume element. The integral is evaluated
at the time when the volume integral of unbound matter reaches its
maximum (compare \fref{fig:unbound_measures}). From this, we define
the average escape velocity $v_\mathrm{esc} = \sqrt{1-W_\infty^{-2}}$.
The results (except for run \texttt{UM\_B0}, which did not contain
this measure) are given in \tref{tab:outcome}.

For the unequal mass model, we found no influence of the magnetic 
field. For the low mass models, which does eject only a small amount of
matter in the first place, we observe a difference comparable to the 
numerical error. We also did not expect a significant impact, since the 
magnetic pressure is always less than $10^{-2}$ of the fluid pressure, 
and even smaller at radii $>50 \usk\kilo\meter$ (see \fref{fig:MHD_ratio}).
We remark that on longer timescales ${\gg}100\usk\milli\second$, baryon 
pollution in the environment surrounding a SMNS can be dominated 
by magnetically induced winds (e.g. \cite{Siegel:2014:6}). 
Moreover, on such timescales neutrino induced winds can also contribute 
significantly (e.g. \cite{Dessart:2009:1681}). 
We also note that the amount of ejected matter might change when 
considering initial NSs with spin, as shown in \cite{Kastaun:2015:064027} 
for different equal mass models.

\section{Summary and Conclusion}
\label{sec:conclusions}

We have presented a set of fully GRMHD simulations of BNS mergers employing the APR4
EOS, which include in particular the first GRMHD evolution of
an unequal-mass BNS with a piecewise polytropic EOS. This set is meant
to cover different scenarios for SGRB central engines: the ``standard'' one in which
a BH is promptly formed (our HM case), and the ``time-reversal'' one
\cite{Ciolfi:2015:36} where a long-lived supramassive NS is the end
result of the merger (LM and UM models). For all simulations, we have
provided a detailed description
of the dynamics, the magnetic field evolution, the ejected matter,
the post-merger remnant properties, and the GW signals.

Both our UM and LM models produce massive disks orbiting the SMNS
remnant, while in the HM model the mass left in the
disk is negligible. Magnetic fields also have an impact on the disk
mass, reducing it by ${\sim}31\%$. 
All our SMNSs exhibit rotation profiles with a slowly rotating core,
outer layers close to Kepler velocity, and a maximum in-between.
This is similar to what was already observed in previous
simulations~\cite{Kastaun:2015:064027} for equal mass systems 
with LS220 and NL3 EOSs. For the first time, we repeated the same detailed 
analysis of the rotation profile also for an unequal mass model,
finding analogous qualitative results. Our findings suggest that our
SMNSs are not supported against immediate collapse by the rotation of
the core, but mainly by the centrifugal support of the outer
layers. This can have crucial implications for the lifetime of the SMNS
and for the possibility of forming a remnant disk after its eventual
collapse. 

In our simulations, magnetic fields do not grow stronger than
$\sim10^{14}~\mathrm{G}$ and the ratio of magnetic to gas pressure
remains always lower than ${\sim}10^{-2}$.  
As a consequence, the impact on the dynamics is only marginal. 
Interestingly, the magnetic field in the SMNS (LM and UM
cases) is strongly toroidal, while outside the post-merger remnant,
both the toroidal and poloidal components have similar strengths. We
also note that the UM case results in a larger amount of baryon pollution 
around the SMNS spin axis. The level of baryon pollution found in both
the UM and LM case can easily choke the formation of a
possible relativistic jet. This suggests that these systems are unlikely
to act as SGRB central engines, unless the late-time collapse of the
SMNS generates the conditions to launch a relativistic jet, as
envisaged in the ``time-reversal'' scenario \cite{Ciolfi:2015:36}.
We remind the reader that our resolutions are not
sufficiently high to accurately resolve the magnetic field
amplification during the merger~\cite{Kiuchi:2015:1509.09205}. Future
simulations employing higher resolution and/or our
subgrid model~\cite{Giacomazzo:2015} will be necessary to shed light on
the impact of magnetic fields on the post-merger dynamics and
possible jet formation. 

For all models we also computed the GW signals. Not surprisingly, the
magnetic field does not affect the GW signal during inspiral, but it has
some small effects in the post-merger signal. 
Higher resolutions might result in stronger magnetic field effects.
Placing our BNS mergers at a distance of 100~Mpc, we find that both advanced Virgo
and advanced LIGO would have no difficulty in detecting the inspiral
and merger GW signal, while the post-merger signal of the LM and UM models 
would be barely visible.
The detection of the post-merger GW signal would have crucial implications
for the NS EOS, as it will clearly indicate the presence of an NS after
merger and the frequency of the main post-merger peak can be used to
constrain the EOS.

SGRBs are not the only EM counterparts that can be expected from these mergers. 
In the case of a SMNS remnant, the GW signal will be very likely 
accompanied/followed by a strong isotropic emission in the X-ray band,
powered by the spindown of the NS \cite{Siegel:2016a,Siegel:2016b}. 
Moreover, our simulations of the unequal mass case show a large amount
of unbound matter, around $0.01 \usk M_\odot$, which could give rise
to macronova emission. Some of this matter will also fall back at
later times and hence provide a possible EM re-brightening of the
central source. 

Note that we use a simplified treatment of thermal effects instead
of a finite temperature nuclear physics EOS (see \cite{Bauswein:2010:84043}
for a discussion of the accuracy of this approach) and that neutrino
radiation is not included. Both will probably have an impact on the 
post-merger phase of the simulation. We consider this work as an 
intermediate step towards the implementation of a full description in 
our code and a useful basis for comparison.

When a post-merger GW signal will be detected with sufficient signal-to-noise 
ratio to extract the main frequency, or even the evolution of this frequency and 
the amplitude, numerical relativity simulations
such as the ones presented in this paper will be crucial to interpret
the findings and draw conclusions on the EOS. The same applies for a 
possible simultaneous detection of EM counterparts. Further, the availability 
of simulated GW signals will be beneficial for the development of better 
GW data analysis tools targeting BNS mergers. For these reasons, we will make
all our gravitational waveforms publicly available, together with
movies (available at \url{stacks.iop.org/CQG/33/164001/mmedia}) visualizing the merger.

\ack
We thank Luca Baiotti for useful discussions and comments. We also
acknowledge support from MIUR FIR grant No.~RBFR13QJYF. Numerical
simulations were run on the cluster Stampede (TACC, USA) via XSEDE
(allocation TG-PHY110027) which is supported by NSF grant
No.~OCI-1053575, on the clusters Fermi and Galileo at CINECA (Bologna,
Italy) via INFN teongrav allocation and via ISCRA grants IsC34\_HMBNS
and IsB11\_MagBNS, and on the cluster Datura at the Albert Einstein
Institute (Potsdam, Germany). We also acknowledge PRACE for awarding
us access to SuperMUC based in Germany at LRZ (grant GRSimStar). B.G.
and T.K. also acknowledge partial support from ``NewCompStar", COST
Action MP1304.

%
%
\appendix

\section{Covergence Test}
\label{sec:app}

\begin{figure}[!ht]
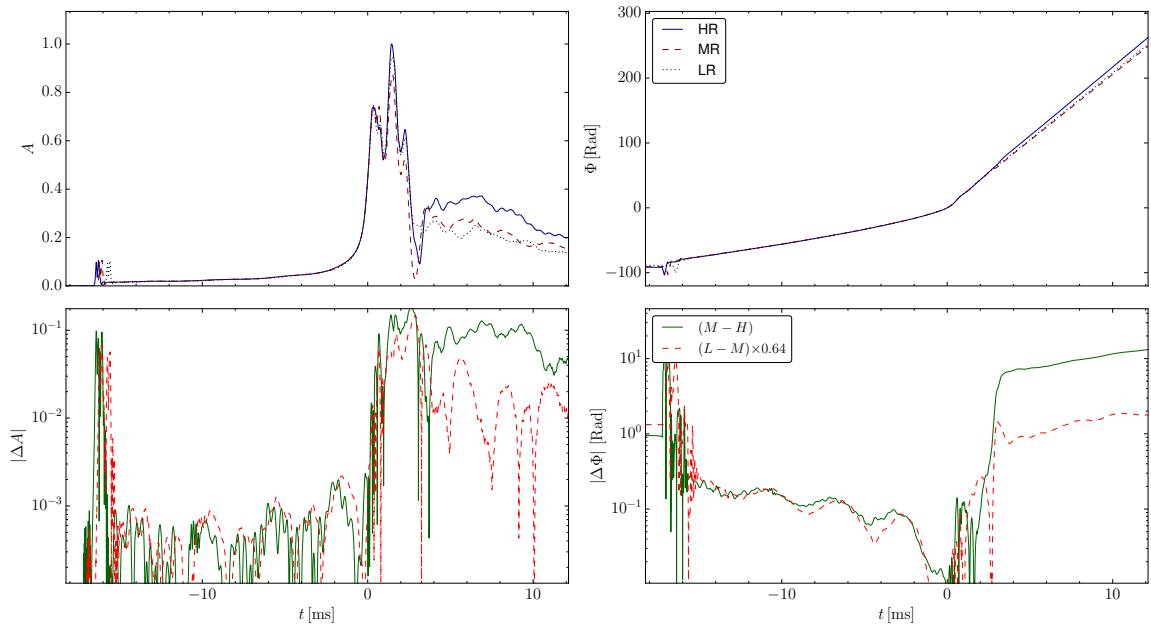

  \includegraphics[width=0.99\textwidth]{{{convergence_gw}}}
  \caption{Comparison of GW signal obtained at resolutions $dx=177.2,
    221.5, 276.9\usk\meter$.  The top left panel shows the amplitude
    of the Weyl scalar $\Psi_4$, normalized to the maximum for the
    highest resolution. The time for each run is shifted such that
    the curves are aligned at time $t_\mathrm{merger}$, the time when
    the strain reaches its maximum.  The top right panel shows the
    (continuous) phase of $\Psi_4$, relative to the phase at
    $t_\mathrm{merge}$.  The bottom panels show the differences
    between consecutive resolutions. The low-medium
    residual has been scaled by a factor 0.64 that corresponds to 
    the second order convergence we expect during inspiral.}
  \label{fig:converge_gw}
\end{figure}

In order to estimate the accuracy of the numerical results, we evolved
model \texttt{UM\_B13} up to $15\usk\milli\second$ after merger using three 
resolutions differing by a factor $1.25$. The intermediate (``medium") 
resolution $dx=221.5\usk\meter$ is the one employed in all the other 
simulations in this work.
\Fref{fig:converge_gw} shows a comparison of the resulting GW signal.
During inspiral, the errors converge, with an overall order around two,
as expected from our scheme. During the merger, the convergence order 
drops to around one. We recall that the numerical scheme is only first 
order accurate when shocks are involved. 
The sharp raise around $t=3\usk\milli\second$ by $2\pi$ in the
difference of the continuous phases between medium and high
resolution has a trivial explanation: the GW signal has a knot in 
the amplitude at this time, during which the phase is irrelevant, 
but very sensitive to errors. However,
even when correcting for the phase jump, we cannot demonstrate 
convergence (of any order) after this point. Instead, we find a 
higher frequency and a much larger amplitude of the late GW signal 
when using high resolution compared to the low and medium resolution.

\begin{figure}
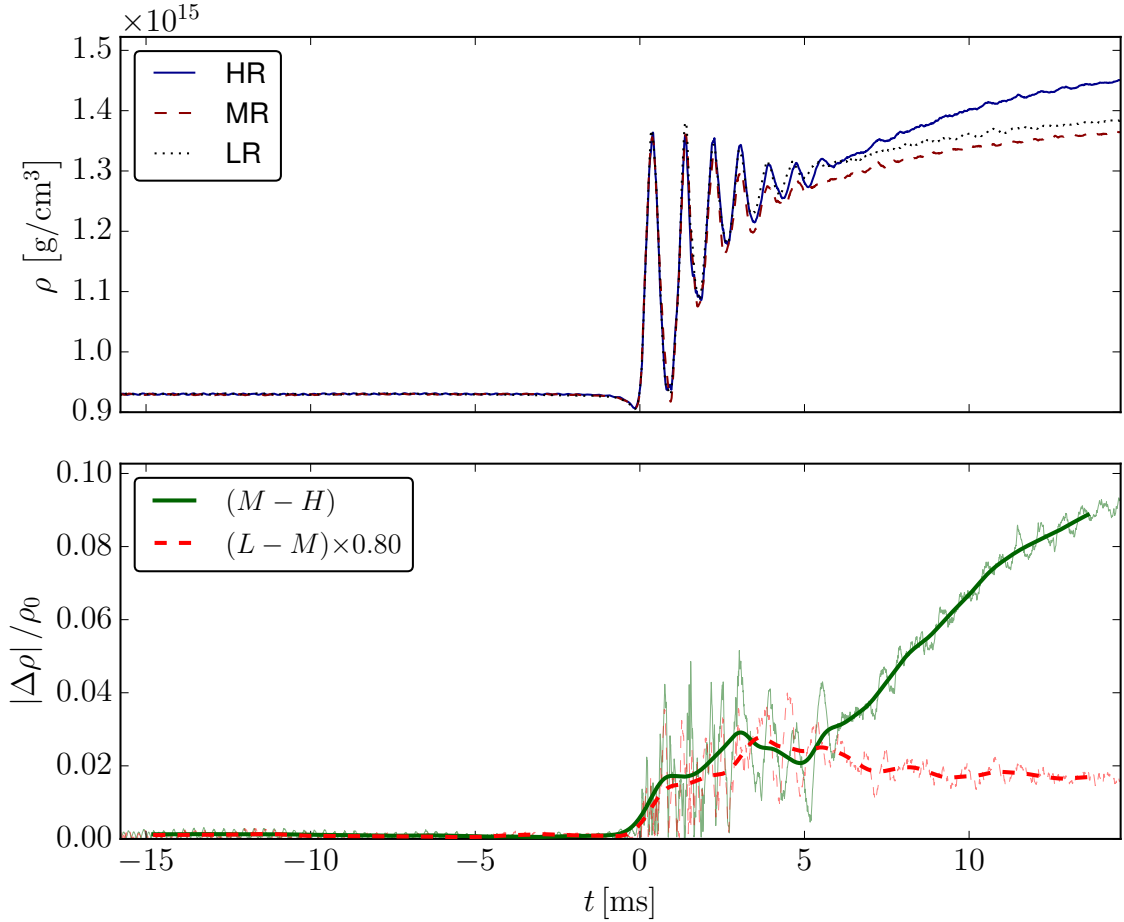

  \centering
  \includegraphics[width=0.99\linewidth]{{{conv_rhomax}}}
  \caption{Upper panel: Evolution of the maximum rest mass density for
    the magnetized unequal-mass model. The different curves
    correspond to low, medium and high resolution. The time coordinate 
    for each curve is relative to the time of the merger.
    Lower panel: Difference of maximum rest-mass density between 
    consecutive resolutions, in units of initial maximum density (thin 
    lines). As a measure of overall convergence, we also plot smoothed 
    versions (thick lines). The low-medium residual was scaled by a 
    factor $0.8$ that corresponds to the first order convergence we expect 
    during merger.}
  \label{fig:rhomax_conv}
\end{figure}

Before offering an explanation, we turn to the convergence of the maximum 
density shown in \fref{fig:rhomax_conv}. During the inspiral, the density 
agrees very well. The differences that appear during merger are on average 
compatible with a convergence order of one. After $5\usk\milli\second$ 
however, we start loosing convergence, and the density for the high-resolution 
case continuously increases with respect to the lower resolutions. What 
causes this worrisome behavior? First, we note that the increase in central
density seems to be a consequence of the decrease in angular momentum due 
to GW radiation. We already saw that the late GW amplitude for the 
high resolution run is larger. Compared to the medium resolution, the 
difference in radiated angular momentum is 
$\Delta J \approx 0.08\usk M_\text{ADM}^2$, which is enough to explain the 
difference in maximum density. The late time evolution more than 
$5\usk\milli\second$ after merger can thus be explained by the differences of 
the remnant's $l=m=2$ oscillation mode amplitude which are already present 
a few $\milli\second$ after merger. 
We recall that the rotation profile undergoes some rapid rearrangement 
shortly after merger, which might be linked also to the KH instability.
It seems very plausible that this also affects the oscillation amplitude. 
We therefore believe that the loss of convergence is mainly due to 
insufficient resolution during the short rearrangement phase, while the
resolution at later times is sufficient. Given a fixed amount of computational 
resources, it might thus be beneficial to spend a larger fraction on this 
phase in order to reduce the overall error.
In any case, it seems likely that the late evolution is as sensitive to 
small physical changes in the binary parameters as it is to the numerical 
error.

\begin{figure}[t]
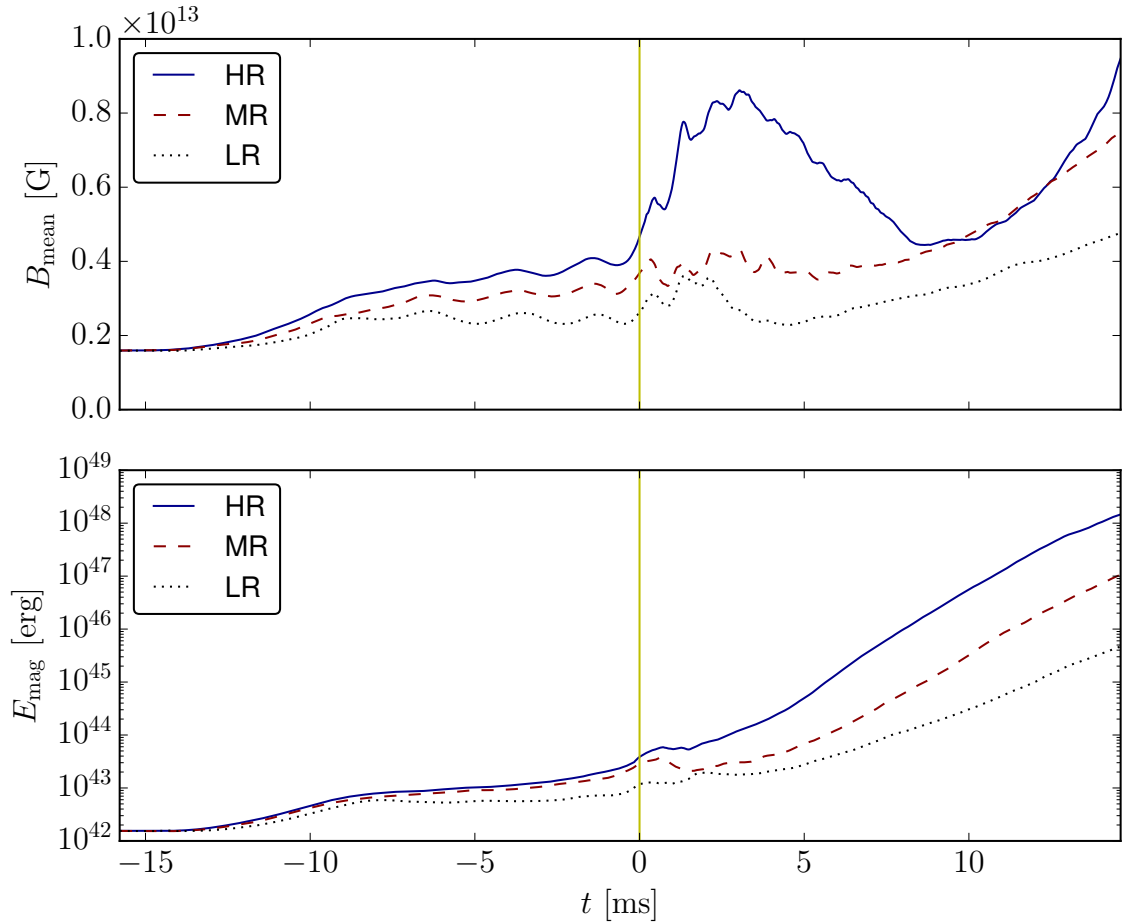

  \centering
  \includegraphics[width=0.99\linewidth]{{{conv_mag}}}
  \caption{Evolution of the density-weighted average of the magnetic
    field strength $B_\mathrm{mean}$ (upper panel) and of the magnetic
    energy $E_\mathrm{mag}$ (lower panel) for the magnetized
    unequal-mass model. The different curves correspond to low,
    medium and high resolution. The vertical line marks the time of
    merger $t=0$.}
  \label{fig:B_conv}
\end{figure}

The mass of ejected matter obtained from the low-, standard-, and
high-resolution runs is 0.0102, 0.0100, and 0.0090, respectively.
Thus, we cannot demonstrate convergence for the ejecta mass. To
estimate the error, we have to blame the lowest resolution. Making
the optimistic assumption that we have first order convergence (as
expected from our scheme in the presence of shock waves) starting from
the standard resolution, we obtain an error of ${\approx}50\%$ for the
ejecta masses obtained at the standard resolutions.

The evolution of the mean magnetic field strength $B_\mathrm{mean}$
(cf. \sref{sec:mag}) and the magnetic energy $E_\mathrm{mag}$ is
shown in \fref{fig:B_conv}. Convergent behavior is observed only up to
the time of merger. In the post-merger phase, higher resolutions exhibit
a much stronger magnetic field amplification. This agrees with the
expectation (cf. \sref{sec:mag}) that the important contribution to the 
amplification given by the KH instability, and possibly the MRI, act on
length scales much too small to be fully resolved in our simulations.  
As a consequence, our results on the evolution of the
magnetic field should be regarded as qualitative. In particular, the
final level of magnetization reached as the system approaches a
quasi-stationary state has to be considered as a lower limit. A more
quantitative investigation, based on higher resolution simulations
and/or on the use of a subgrid model \cite{Giacomazzo:2015} will be
the subject of future work.

\begin{figure}[t]
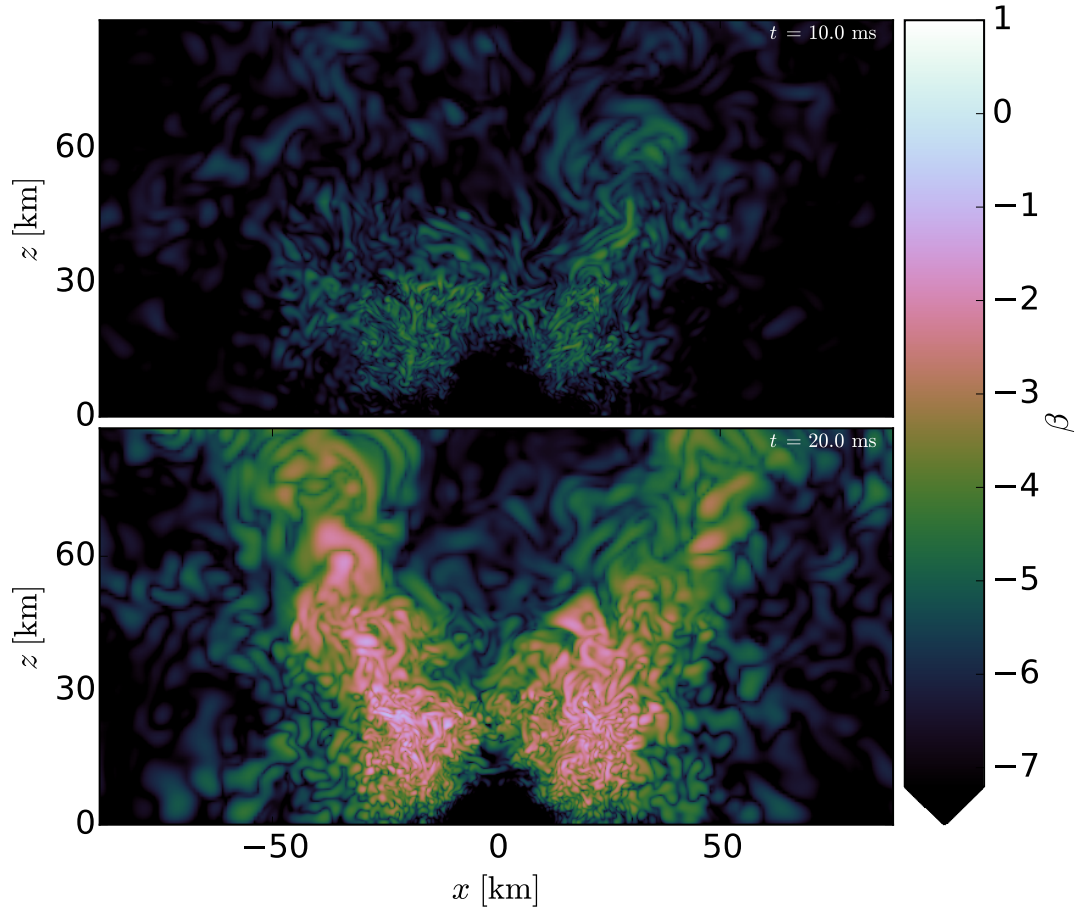

  \centering
  \includegraphics[width=0.99\linewidth]{{{2Dxz_MHD-ratio}}}
  \caption{Meridional view of the magnetic-to-fluid pressure 
  ratio (see text) for the magnetized unequal-mass model at $t=10$ 
  and $20\usk\milli\second$.} 
  \label{fig:MHD_ratio}
\end{figure}

\section{Fluid pressure domination over magnetic field}
\label{sec:app_mag}

In \fref{fig:MHD_ratio} we show the ratio of magnetic pressure
over fluid pressure in the meridional plane for the unequal-mass case,
10 and 20~ms after merger. The ratio is defined as $\beta\equiv
b^2/2p$, where $b^2\equiv b^\mu b_\mu$ and $b^\mu$ is the 4-vector of
the magnetic field as measured by the comoving observer
\cite{Giacomazzo:2007:235}.  We find that the maximum $\beta$ is
achieved in the torus and grows up to $\sim\!\mathrm{few}\times
10^{-2}$. The growth of $\beta$ is significantly slower towards the
end of the simulation, following the behavior of the magnetic field
amplification. We conclude that magnetic fields remain dynamically
subdominant at all times and everywhere in the shown domain.  The same
conclusion applies to the other magnetized cases considered in this
work.

\section*{References}
\bibliographystyle{iopart-num} 
\bibliography{trento}

\end{document}